\documentclass[10pt,twocolumn,letterpaper]{article}

\usepackage{cvpr}              

\usepackage{graphicx}
\usepackage{amsmath}
\usepackage{amssymb}
\usepackage{booktabs}

\usepackage{graphicx}
\usepackage{amsmath}
\usepackage{amssymb}
\usepackage{booktabs}
\usepackage{color}
\usepackage{multirow}
\usepackage{balance}

\usepackage[pagebackref,breaklinks,colorlinks]{hyperref}

\usepackage[capitalize]{cleveref}
\crefname{section}{Sec.}{Secs.}
\Crefname{section}{Section}{Sections}
\Crefname{table}{Table}{Tables}
\crefname{table}{Tab.}{Tabs.}

\def\cvprPaperID{8914} 
\def\confName{CVPR}
\def\confYear{2023}

\begin{document}

\title{Super-High-Fidelity Image Compression via Hierarchical-ROI and Adaptive Quantization}

\author{Jixiang Luo\\
Sensetime Research\\
{\tt\small jixiangluo85@gmail.com}
\and
Yan Wang\\
Tsinghua University\\
{\tt\small wangyan@air.tsinghua.edu.cn}
\and
Hongwei Qin\\
Sensetime Research\\
{\tt\small qinhongwei@sensetime.com}
}
\maketitle

\begin{abstract}
Learned Image Compression (LIC) has achieved dramatic progress regarding objective and subjective metrics. MSE-based models aim to improve objective metrics while generative models are leveraged to improve visual quality measured by subjective metrics. However, they all suffer from blurring or deformation at low bit rates, especially at below $0.2bpp$. Besides, deformation on human faces and text is unacceptable for visual quality assessment, and the problem becomes more prominent on small faces and text. To solve this problem, we combine the advantage of MSE-based models and generative models by utilizing region of interest (ROI). We propose Hierarchical-ROI (H-ROI), to split images into several foreground regions and one background region to improve the reconstruction of regions containing faces, text, and complex textures. Further, we propose adaptive quantization by non-linear mapping within the channel dimension to constrain the bit rate while maintaining the visual quality. Exhaustive experiments demonstrate that our methods achieve better visual quality on small faces and text with lower bit rates, e.g., \textbf{0.7X} bits of HiFiC~\footnote{https://hific.github.io/} and \textbf{0.5X} bits of BPG.

\end{abstract}

\begin{figure}[tb]
    \centering
    \includegraphics[width=0.9\linewidth]{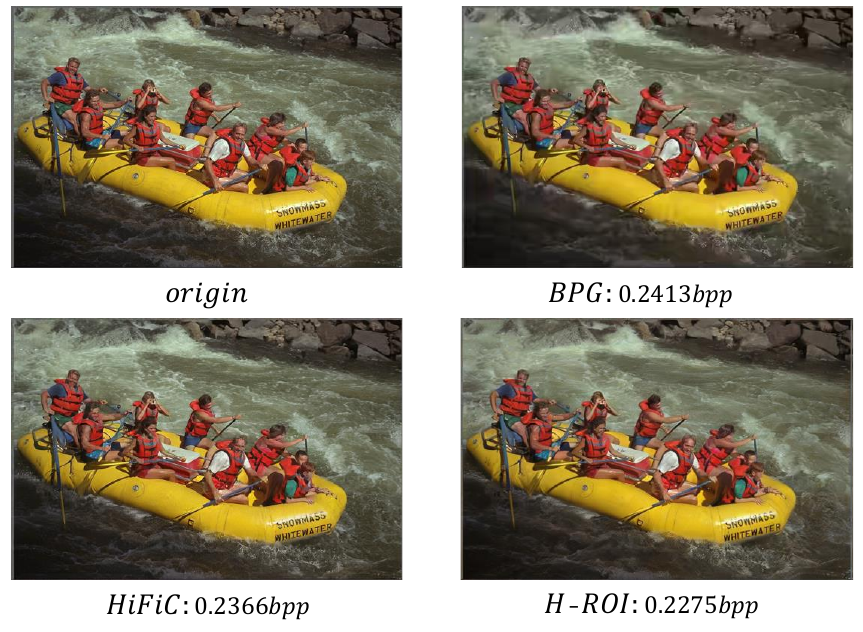}
    \includegraphics[width=0.85\linewidth]{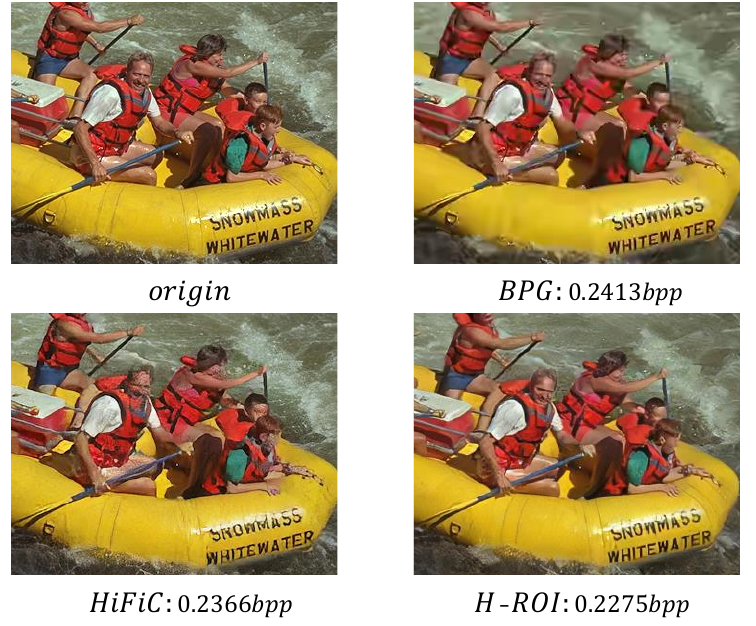}    
    \caption{The visual quality of Kodim14 with H-ROI v.s. BPG and HiFiC. Our method shows higher fidelity for human faces and text on the boat with a smaller bpp.}
    \label{fig:kodim14_01}
\end{figure}

\begin{figure*}[tb]
    \centering
    \includegraphics[width=0.85\linewidth]{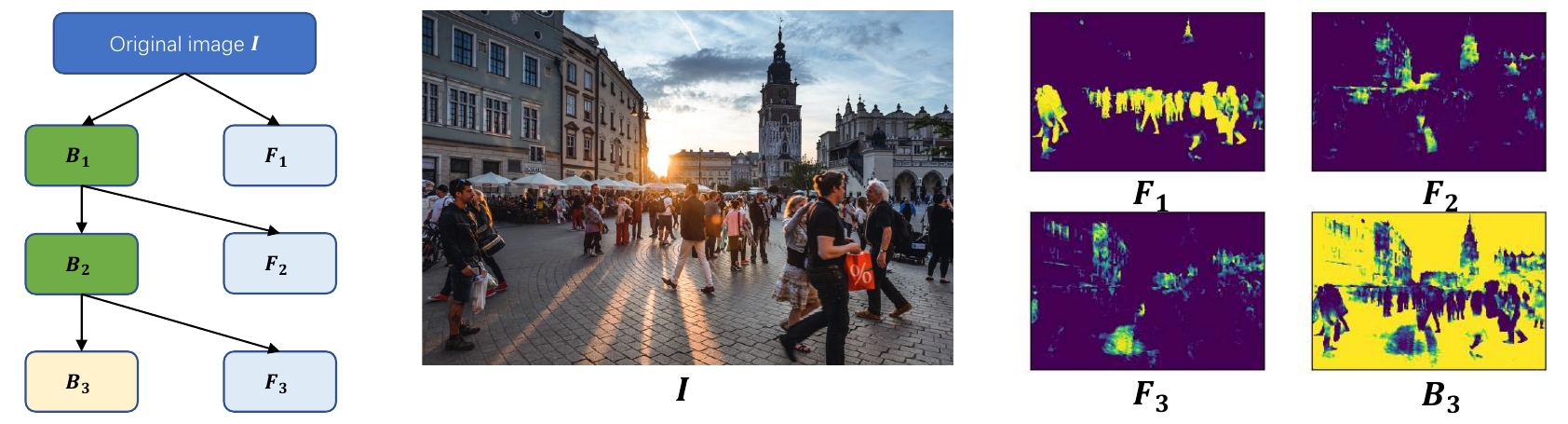}
    \caption{Hierarchical-ROI with a salient object detection network. $I$ is the original image. $F_{i},i=1,2,3, B_{i},i=1,2,3$ represent the foreground and background of the $i_{th}$ layer. The right column is visualization of salient objects in yellow.}
    \label{fig:hier-roi}
\end{figure*}

\section{Introduction}
Learned Image Compression (LIC) with deep neural networks has gone through rapid development, outperforming traditional methods like JPEG~\cite{jpeg} and BPG~\cite{bpg} in terms of objective metrics like Peak Signal-to-Noise Ratio (PSNR) and Multi-scale Structural Similarity (MS-SSIM). The main transformation with hyperprior framework  ~\cite{imagecnn5, imagecnn7,cheng2020learned,gao2021neural,guo2021causal,lee2019end,li2020efficient,lin2020spatial,liu2019non,liu2020unified,ma2021cross,ma2020end,yang2020improving,zhao2021universal,fu2021learned} models image representation with the constraint of entropy by introducing the concept of Variational AutoEncoder (VAE)~\cite{kingma2013auto}, acting as the basis of later works which further improve rate-distortion performance. Besides, context is one of the most important modules to provide a more accurate estimate of the probability of the symbols being encoded by the arithmetic coder. Thus, context models~\cite{imagecnn6,imagecnn7,cheng2020learned,he2021checkerboard, fu2021learned} are proposed, utilizing the causality of latent symbols within spatial and channel dimensions. Even though LIC achieves better performance compared to traditional codecs in terms of PSNR and MS-SSIM, LIC still suffers from compression artifacts similar to JPEG compression noise, which is interpreted as perception-distortion trade-off \cite{blau2018perception}. Patel et al.~\cite{patel2021saliency} proposed deep perceptual compression to align with human eyes using deep perceptual loss. However, adding losses alone is incapable of capturing the characteristics of human eyes. Contour and texture details are still missing at low bit rates. 



To obtain better visual perceptual quality of the reconstruction, previous works introduce generative adversarial network (GAN)\cite{goodfellow2014generative} to enhance perceptual quality by generating more details than those MSE-optimized or MS-SSIM-optimized models. Besides, Agustsson et al.\cite{agustsson2019generative} utilize adversarial training to efficiently compress images at low bit rates to maintain details with high frequency. Meanwhile, HiFiC~\cite{mentzer2020high} introduces a generator and a conditional discriminator with latents for perceptual quality, resulting in reconstruction more consistent with human eyes. However, they all suffer from common problems caused by GAN, such as unnatural texture, drifted color and some new content from generated noise. Among these phenomena, human faces and text are more sensitive to deformation, especially at low bit rates where small deformation can result in extremely poor visual quality.  

Region of Intrest (ROI) leverages the importance of image content to allocate bits, assigning enough bits to sophisticated textures to maintain high quality while allocating just a few bits to smooth regions. There exist many excellent works for salient object or region detection~\cite{chen2020global, cheng2014global,hou2017deeply}, which can mark out people, moving objects and other objects with rich colors. Meanwhile, compression with ROI can be categorized into masking on the original image and masking on latents. The works of \cite{cai2019end, patel2019deep} first utilize a saliency detection network to generate ROI mask, combine it with latents to obtain more efficient image representation, and adjust the quantization step of latents to maintain the most important features. Besides, some methods~\cite{chang2021thousand, chang2022consistency} compress the salient mask using lossless methods along with the latent bitstream to reconstruct facial scenarios at extremely low bit rates, like those below $0.3bpp$.
Besides masking latents, Ma et al.\cite{ma2021variable} also introduce a background loss and an ROI loss for different image regions. However, all these methods increase computational complexity within  the compression process
because of additional ROI extracting network. 

In this paper, our contribution is threefold:
\begin{itemize}
    \item We can balance extremely low bit rate and high visual quality, dramatically improving the reconstruction of human faces and text in the foreground, especially for small faces and text. Besides, we further maintain the realism of structure in the background. 
    \item We utilize hierarchical ROI (H-ROI) to split the image into several foreground regions and one background region. Then we apply perceptual loss and GAN loss for the background and apply MSE loss for foregrounds with different importance factors.
    \item We decouple the adaptive quantization from the ROI mask to make our inference process more efficient by adjusting the quantization boundary with non-linear transformation, which further reduces the bit cost of the background while maintaining visual quality. 
\end{itemize}

\begin{figure*}[tb]
    \centering
    \includegraphics[width=0.85\linewidth]{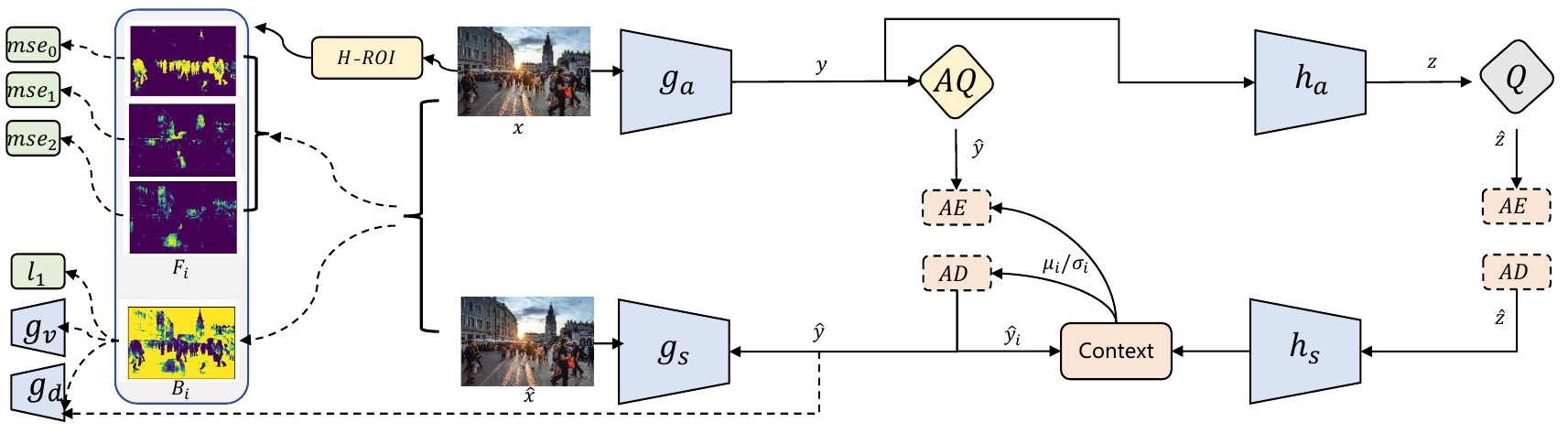}
    \caption{Diagram of the network adopted. The right part is ELIC~\cite{elic}. We use the same architecture for $g_a, g_s, h_a$ and $h_s$ as the original paper. \textit{Context} denotes the spatial-channel context model described in ELIC. \textit{Q, AQ} are the quantization and adaptive quantization. \textit{AE, AD} are the arithmetic encoding and decoding. The left part shows the adversarial training $g_d$, which has the same discriminator structure as HiFiC~\cite{mentzer2020high}, and perceptual learning $g_v$ which we train with VGG network~\cite{simonyan2014very} and $l_1$ loss. We use MSE loss $mse_{i}, i=0,1,2$ for foregrounds at different levels. }
    \label{fig:elic-arch}
\end{figure*}

\section{Background}
\subsection{LIC with Context Models}
Image compression task aims to optimize the rate-distortion function $\mathcal{R} + \lambda \mathcal{D}$, where $\mathcal{R}$ is the bit rate, $\mathcal{D}$ is the distortion, and $\lambda$ is the Lagrange multiplication factor. Denoting the image as $x$, encoder and decoder of the neural network as $g_a$ and $g_s$, the overall loss function is written as follows:
\begin{equation}
\mathcal{L}= \mathbb{E}[-\log p(g_a(x)) + \lambda d(x, g_s(g_a(x)))]     
\label{rd_loss}
\end{equation}
where $\mathbb{E}$ is the expectation over $p(x)$, $g_a$ extracts the input image $x$ as latent variable $y = g_a(x)$ and $g_s$ transforms it into reconstructed image $\hat{x}$.

To eliminate the statistically redundant information, the auto-regressive context model is introduced to promote compression performance by leveraging the causality of latents and conceptual similarity. To be specific, the estimation of current symbol $y_{i,j,k}$ can leverage previous symbols $y_{<i, <j,<k}$:
\begin{equation}
     p(y_{i,j,k}|y_{<i,<j,<k}) = p(y_{i,j,k}|\Psi(y_{<i,<j,<k}))
\end{equation}
where $i$ represents the channel dimension and $j,k$ refer to the spatial dimensions. $\Psi$ can be various in combinations of channel or spatial dimensions to construct the context model. Minnen~\etal~\cite{imagecnn7} utilizes spatial information while ~\cite{minnen2020channel, elic} use channel context modeling.

\subsection{LIC with Generative Adversarial Networks}
Some works ~\cite{mentzer2020high, Chen2021MCM, Gao2021clic-huawei,santurkar2018generative,yang2020learned} take 
image restoration task as image generation, and 
GAN has been a powerful tool for image generation or style transfer. Meanwhile, $g_s$ can be regarded as a generator, while discriminator $g_d$ is introduced to eliminate the discrepancy between objective metrics and the human visual system.  
Different from the training of GAN in which generator and discriminator update their weight alternately, $g_s, g_a$ and $g_d$ in the image compression framework with discriminator are jointly trained~\cite{mentzer2020high}. To train the discriminator $g_d$, an auxiliary discriminator loss like binary cross-entropy  is introduced:
\begin{equation}
    \mathcal{L}_{gan} = -\mathbb{E}\left[\log g_d(x, \hat y)\right] - \mathbb{E}\left[\log \left(1-g_d(\hat x, \hat y)\right)\right],
    \label{eq:gan-bce-D}
\end{equation}
where $\hat y$ acts as a bridge between the original image and the reconstructed image, improving the visual quality.

\subsection{LIC with Salient Object Detection}
Spatial and temporal information is essential for salient object detection as it facilitates the detection of human attention. It leverages the binary cross entropy loss between ground truth and prediction :
\begin{equation}
    \mathcal{L}_{bce} = -\frac{1}{N}[m_i \log(p(m_i)) + (1-m_i) \log(1 - p(m_i))],
\end{equation}
where $N$ is the total number of pixels, $m_i$ is the ground truth, $p(m_i)$ is the predicted probability. 

Then mask $m$ is applied to the latents generated from $g_a(x)$ as described in~\cite{cai2019end, patel2019deep,ma2021variable} to modify Eq.~\ref{rd_loss}:
\begin{equation}
\mathcal{L}= \mathbb{E}[-\log p(m\otimes g_a(x)) + \lambda d(x, g_s(m\otimes g_a(x)))],     \label{eq:roi-latent}
\end{equation}
where $\otimes$ is the element-wise operator. The inference process must balance the complexity of salient object detection and rate-distortion performance since the mask $m$ and latents $g_a(x)$ are related in the optimization. Gu \etal ~\cite{gu2020PCSA} proposed PCSA which leverages Pyramid structure and Constrained Self-Attention to capture salient objects with various scales. Besides, it can save computation and memory costs by looking at neighbor regions instead of global areas.

\section{Architecture}\label{architecture}

We adopt ELIC~\cite{elic} framework as our coding architecture. Fig.~\ref{fig:elic-arch} shows its diagram, which consists of $g_a, g_s, h_a, h_s$. When optimizing MSE, it achieves better RD performance than VVC~\cite{bross2021overview} w.r.t. both PSNR and MS-SSIM. \textit{Q, AQ} are the quantization and adaptive quantization. \textit{Context} is the probability engine for \textit{AE, AD}, which are the arithmetic encoder and decoder. $g_d$ is the discriminator conditioned on latents and $y$, and $g_v$ is the VGG~\cite{simonyan2014very} network we utilize for LPIPS\cite{zhang2018unreasonable} loss. H-ROI is described in Fig.~\ref{fig:hier-roi}. Then, we have the following training loss for the first stage:
\begin{equation}
    \mathcal{L}_{stage1} = \mathcal{R} + \lambda_0 * MSE,
\end{equation}
where $R, MSE$ are the bit rate and mean square error function. For the second stage, we have the following loss formulation:
\begin{align}
    \mathcal{L}_{stage2} & = \mathcal{R} + \sum^2_{i=0} [\lambda_i * MSE_i \otimes m_i] + [\lambda_{lpips} * \mathcal{L}_{lpips} \notag \\
    & + \lambda_{l1} * \mathcal{L}_1 + \lambda_{gan} * \mathcal{L}_{gan}] \otimes (1 - \sum^2_{i=0} m_i) ,
\end{align}
where $m_{i}, i=0,1,2$ are the mask generated as shown in Fig.~\ref{fig:hier-roi} for foregrounds, $\lambda_{i}, i=0,1,2, \lambda_{lpips, gan, l1}$ are the Lagrange multiplers, and we set $\lambda_0 \geq \lambda_1 \geq \lambda_2$. $\mathcal{L}_{lpips, gan, 1}$ are the LPIPS~\cite{zhang2018unreasonable}, adversarial and $l_1$ norm loss function. Besides, $1 - \sum^2_{i=0}m_i$ represents the mask for the background. 

\begin{figure}
    \centering
    \includegraphics[width=0.85\linewidth]{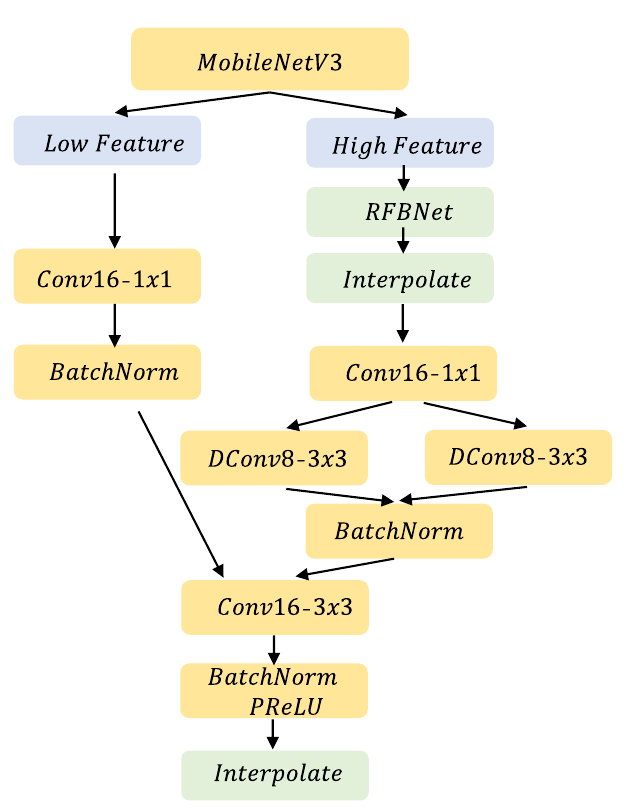}
    \caption{PCSA is simplified from~\cite{gu2020PCSA} in H-ROI. MobileNetV3~\cite{howard2019searching} is used to extract low-dimensional and high-dimensional features. \textit{Conv16-1x1} represents the convolutional layer with $1 \times 1$ kernel and $16$ output channels, while \textit{DConv8-3x3} denotes the dilated convolutional layer with dilation $3$ and $8$ output channels.  \textit{BatchNorm} and \textit{PReLU} are the activation function. \textit{interpolate} denotes the bilinear upsampling. }
    \label{fig:pcsa}
\end{figure}

\begin{figure*}
    \centering
    \includegraphics[width=0.9\linewidth]{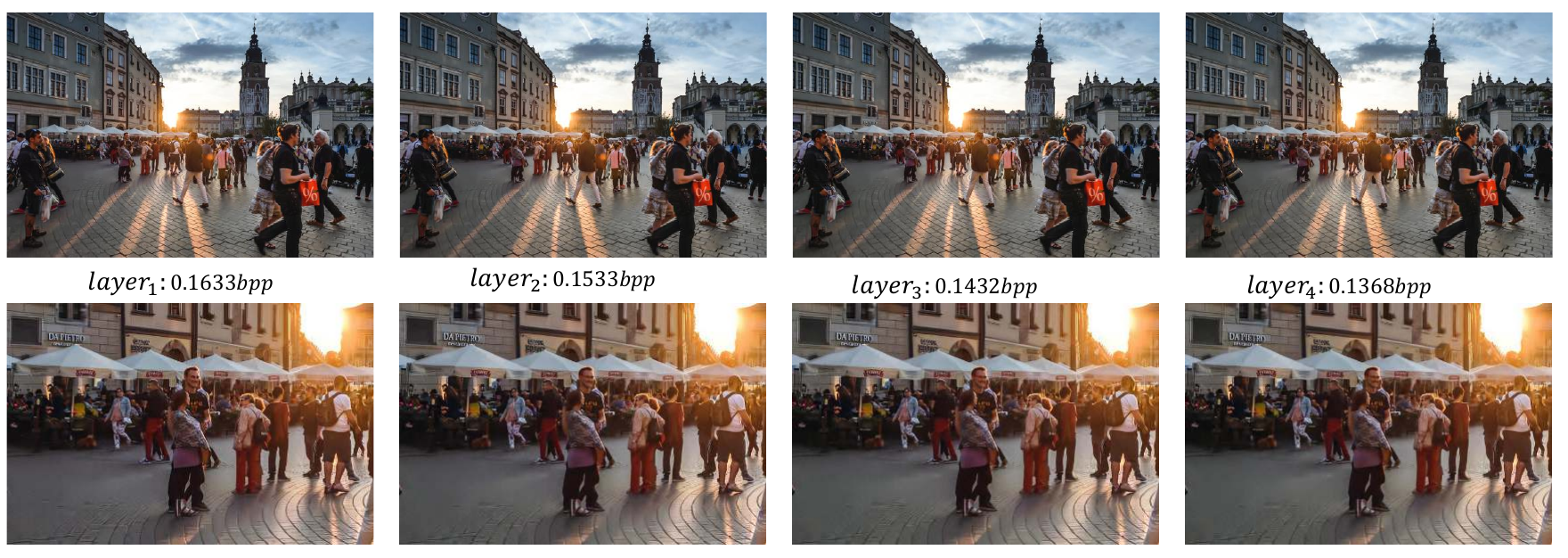}
    \caption{The influence of quantization with different layers. ${layer_1}$ means no adaptive quantization, ${layer_2, layer_3, layer_4}$ means $\epsilon_1, \epsilon_2, \epsilon_3$ are applied for quantization. }
    \label{fig:quant_layer}
\end{figure*}

\section{Hierarchical-ROI}
We apply the salient detection network $\psi$ on original image $I$ to obtain foregrounds $F_{i},i=1,2,3$ and backgrounds $B_{i},i=1,2,3$:
\begin{align}
    & F_1, B_1 = \psi(I) \notag \\  
    & F_2, B_2 = \psi(B_1) \notag \\ 
    & F_3, B_3 = \psi(B_2) 
\end{align}
To obtain the hierarchical attention, we feed the original image into $\psi$ to get the first foreground $F_1$ and background $B_1$. Then, we take $B_1$ as input and feed it into the same neural network $\psi$ to get $F_2, B_2$. The detailed process is shown in Fig~\ref{fig:hier-roi}. The first foreground consists of people on the street which attract the most attention, while the second and third foregrounds include the tower and sunshine, which are more attractive compared to the background. 

Besides, as shown in Fig.~\ref{fig:pcsa} we utilize MobileNetV3~\cite{howard2019searching} as the backbone to accelerate the training process. More specifically, we only use one pre-trained salient detection network to complete all region detection. We obtain the low-dimensional feature from shallow layers of MobileNetV3 and high-dimensional feature from deeper layers. Then, RFBNet~\cite{liu2018receptive} is used to process high-dimensional features to enhance the accuracy of detection. Bilinear interpolation is applied since the resolution of high-dimensional features is of smaller size than low-dimensional features. At last, the low-dimensional and the high-dimensional features after going through several convolutional layers, are concatenated and interpolated to match the resolution of the input. 

\begin{figure}
    \centering
    \includegraphics[width=0.95\linewidth]{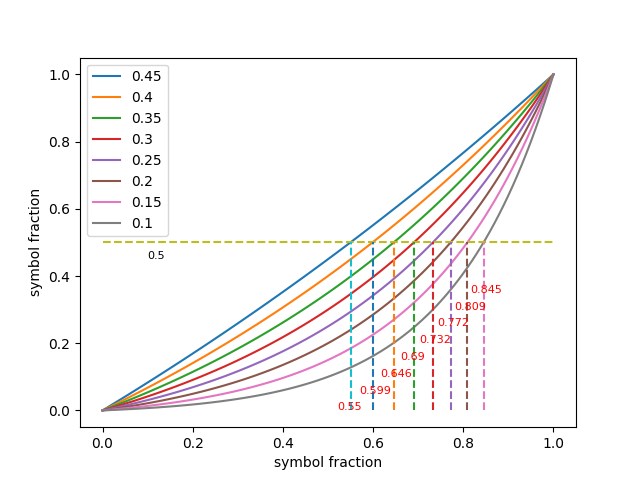}
    \caption{Non-linear mapping. X/Y-axis is the fraction of latents. We map the larger part of the X-axis into the smaller part of the Y-axis. Thus, the original boundary of uniform quantization, which is $0.5$, has been changed to the red annotations. And the lines of $\epsilon = \{0.45, 0.4, \cdots, 0.1\}$ denote the nonlinearity.}
    \label{fig:non_linear}
\end{figure}

\begin{figure}
    \centering
    \includegraphics[width=0.95\linewidth]{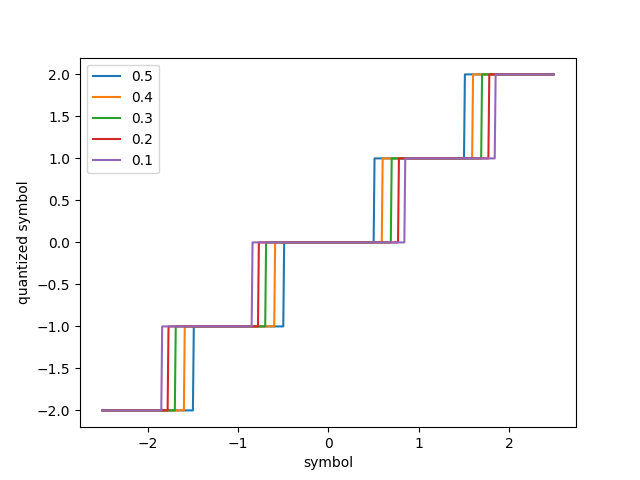}
    \caption{Non-linear quantization. We apply Fig.~\ref{fig:non_linear} to both negative and positive fractions to obtain the centrosymmetric quantization range. The quantized symbol $0$ has a widened range while other symbols have a range $1$ but different quantization boundaries.}
    \label{fig:quant}
\end{figure}

\begin{figure}
    \centering
    \includegraphics[width=0.8\linewidth]{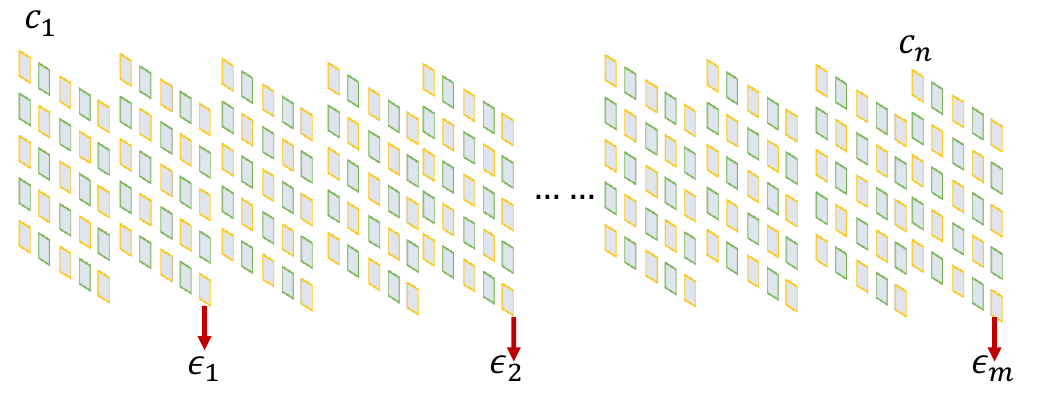}
    \caption{Adaptive quantization with channel groups. $c_1, \cdots, c_n$ denote the channels of latents with total number $n$. For different channel groups, we set $\epsilon_1, \cdots, \epsilon_m$ for $m$ channel groups, where $\epsilon_1 \geq \cdots \geq \epsilon_m$. The \textcolor{red}{red arrows} represent the last channel using $\epsilon_i$.}
    \label{fig:quant_layer2}
\end{figure}

\section{Adaptive Quantization}
LIC usually utilize uniform quantization to quantize the latents and then encode them with arithmetic coding. We adopt the idea of RDOQ~\cite{hoang1996rate} to minimize the magnitude of latents by adjusting the quantization boundary, which further reduces the bit cost of the background while maintaining visual quality.
Besides, we leverage adaptive channel-wise quantization described below. 

To control the bit rate, we minimize the magnitude of latents to constrain its entropy. Then we have the following equation:
\begin{equation}
    \hat{y} = \lfloor \lfloor y \rfloor + \phi(y - \lfloor y \rfloor)  \rceil,
\end{equation}\label{eq: round}
where $\lfloor * \rfloor, \lfloor * \rceil$ represent the floor and the round operator. $\phi$ is the non-linear function to map the fractions into smaller ranges as shown in Fig.~\ref{fig:non_linear}:
\begin{equation}
    \phi(t) = e^{a * t + b} + c,
\end{equation}
where $a,b,c$ are the parameters determined by $\phi(0.0) = 0.0, \phi(0.5) = \epsilon, \phi(1.0) = 1.0$. For simplification, $\epsilon$ is set to $\{0.45, 0.4, \cdots, 0.1\}$. When $\epsilon = 0.0$, adaptive quantization is degenerated to the floor($\lfloor * \rfloor$) function. 

We balance the bit rate and visual quality to avoid the performance loss caused by floor quantization, and then adaptive quantization can adapt the image content as it manipulates each channel of latents independently and each channel of latents represents different image contents, like the details with high frequency or low frequency. 

To be specific, we obtain the non-linear quantization as shown in Fig.~\ref{fig:quant}. $\epsilon$ is chosen from the subset with $\{ 0.4, 0.3, 0.2, 0.1\}$. When $\epsilon$ decreases, the range of values quantized to $0$ increases and the thresholds which decide symbols to be quantized into the nearest integer above its current value become larger.

\begin{figure}
    \centering
    \includegraphics[width=0.49\linewidth]{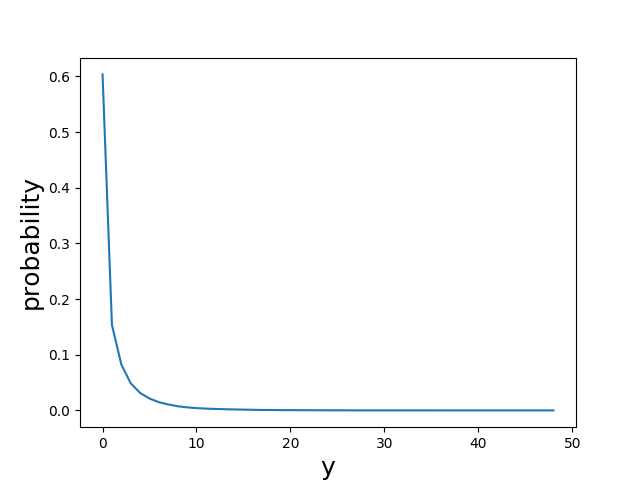}
    \includegraphics[width=0.49\linewidth]{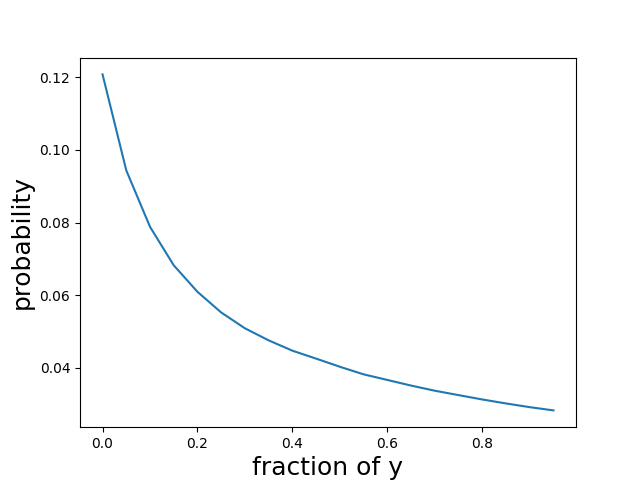}    
    \caption{The left part is the statistical distribution of the absolute value of latents $y$. The right part is the statistical distribution for the fraction $y-\lfloor y \rfloor$ of latents.}
    \label{fig:stat_quant}
\end{figure}

We verify the effectiveness of adaptive quantization using the statistical distribution of latents as shown in Fig.~\ref{fig:stat_quant}, which are collected from the CLIC 2022 testing dataset~\cite{clic2022} at around $0.2bpp$. The left part is the distribution of the absolute value of $y$, mostly lying in the range within $10$ at the low bit rates. Since almost all previous works adopt uniform quantization with the fraction boundary of $0.5$, the right part visualizes the distribution of the fraction of $y-\lfloor y  \rfloor$. We shift the threshold value $0.5$ to the right as shown in Fig.~\ref{fig:non_linear} to $\{0.55, 0.599, \cdots, 0.845\}$. By adjusting the boundary, we control the proportion of latents from which $1$ is subtracted. To some extent, the proportion is decided by the image content, e.g. some smooth regions require fewer bits. Thus, the corresponding latents can be 
modified and we can maintain the visual quality while minimizing the bit rate.  

\begin{table*}[]
\centering
\caption{Quantitative results with PSNR, MS-SSIM and LPIPS for Kodak, CLIC2022 testing dataset and a subset of CrowdHuman testing dataset}\label{tab:quantitative}
\resizebox{\textwidth}{!}{ 
\begin{tabular}{cllllllllllll}
\cline{2-13}
\multicolumn{1}{l}{}                        & \multicolumn{4}{c|}{Kodak}  & \multicolumn{4}{c|}{CLIC2022}  & \multicolumn{4}{c}{CrowdHuman Test30}                
\\ \cline{2-13} 
 & bpp & PSNR & MS-SSIM & \multicolumn{1}{c|}{LPIPS} & bpp & PSNR & MS-SSIM &  \multicolumn{1}{c|}{LPIPS} & bpp & PSNR & MS-SSIM & LPIPS \\ \hline \hline
\multicolumn{1}{c|}{\multirow{3}{*}{BPG}}   & 0.0952                   & 27.0262                   & 0.8908                       & 0.4741                     & 0.0888                   & 27.7474                   & 0.9185                       & 0.4443  & 0.0858 & 26.7938 & 0.9258 & 0.3920                   \\
\multicolumn{1}{c|}{}                       & 0.2145                   & 29.5042                   & 0.9375                       & 0.3769                     & 0.1877                   & 30.0832                   & 0.9495                       & 0.3727  & 0.1730 & 29.6214 & 0.9577 & 0.3005                   \\
\multicolumn{1}{c|}{}                       & 0.4383                   & 32.2964                   & 0.9659                       & 0.2782                     & 0.3604                   & 31.9726                   & 0.9642                       & 0.3085    & 0.3281 &	32.7234 &	0.9756 &	0.2190 
                \\ \hline \hline 
\multicolumn{1}{c|}{\multirow{3}{*}{ELIC}}  & 0.1032                   & 27.1409                   & 0.9033                       & 0.4704                     & 0.1098                   & 27.9678                   & 0.9336                       & 0.4418   &   0.1375 &	27.0942 &	0.9356 &	0.3899 
               \\
\multicolumn{1}{c|}{}                       & 0.3426                   & 32.5643                   & 0.9711                       & 0.2797                     & 0.2109                   & 32.0432                   & 0.9703                       & 0.3320    &    0.2909 &	31.4825 &	0.9738 &	0.2632 
             \\
\multicolumn{1}{c|}{}                       & 0.5213                   & 34.7049                   & 0.9819                       & 0.2267                     & 0.2809                   & 33.0728                   & 0.9762                       & 0.3052   &  0.3854 &	32.6712 &	0.9796 &	0.2331 
                \\ \hline \hline 
\multicolumn{1}{c|}{\multirow{3}{*}{HiFiC}} & 0.1826                   & 27.5647                   & 0.9372                       & 0.2461                     & 0.1518                   & 28.5476                   & 0.9521                       & 0.2338   &     0.1524 &	27.5716 &	0.9555 	& 0.2401 
             \\
\multicolumn{1}{c|}{}                       & 0.3511                   & 29.6538                   & 0.9643                       & 0.1871                     & 0.2900                   & 30.4707                   & 0.9708                       & 0.1837     &  0.2898 &	29.6010 & 0.9740 	& 0.1845 
              \\
\multicolumn{1}{c|}{}                       & 0.5377                   & 31.7428                   & 0.9775                       & 0.1484                     & 0.4410                   & 32.2562                   & 0.9799                       & 0.1552       & 0.4286 &	31.4336 &	0.9826 & 	0.1482 
             \\ \hline \hline
\multicolumn{1}{c|}{\multirow{3}{*}{ours}}  &0.1854 &	27.8794 &	0.9362 
                     & \underline{0.2028}  &   0.1486 &	28.5015 &	0.9510 &	\underline{0.2127} & 0.1951 &	27.6499 &	0.9564 &	\underline{0.2047}    
                     \\
\multicolumn{1}{c|}{}                       & 0.3346                   & 30.8858                   & 0.9645                       & \underline{0.1480}    & 0.2707 &	31.4364 &	0.9705 &	\underline{0.1822} &	 0.3724 &	31.0894 &	0.9766 & 	\underline{0.1467}                      \\
\multicolumn{1}{c|}{}                       & 0.4449                   & 32.0174                   & 0.9737                       & \underline{0.1214}  &   0.3632 &	32.4728 &	0.9774 & 	\underline{0.1327} &	0.4860 	& 32.2559 &	0.9824 &	\underline{0.1226 }
                        \\ \hline \hline 
\end{tabular}
}
\end{table*}

\begin{figure}
    \centering
    \includegraphics[width=0.49\linewidth]{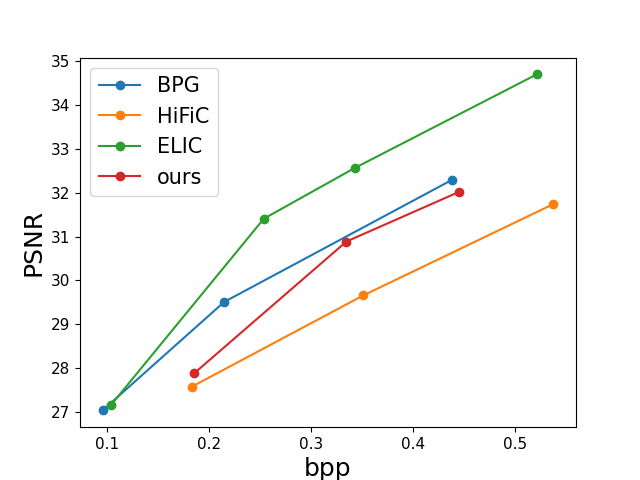}
    \includegraphics[width=0.49\linewidth]{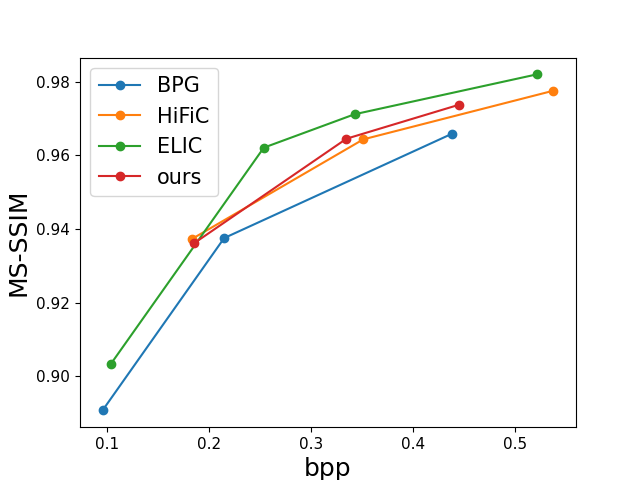}
    
    \includegraphics[width=0.49\linewidth]{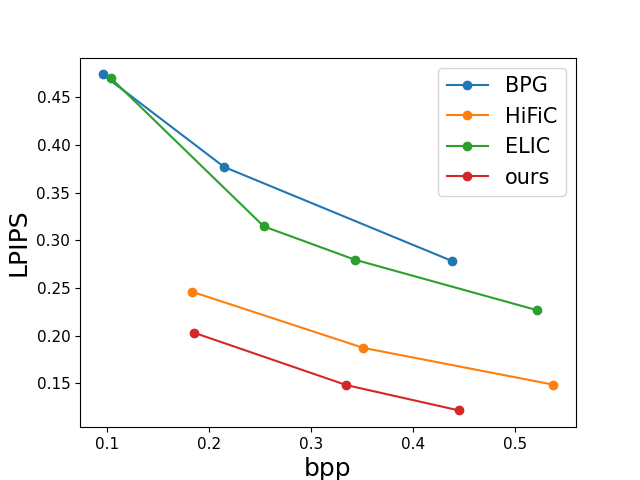}
    \caption{Performance on PSNR, MS-SSIM and LPIPS of our method, HiFiC, ELIC and BPG on Kodak dataset.}
    \label{fig:lpips}
\end{figure}

\begin{figure*}
    \centering
    \includegraphics[width=0.99\linewidth]{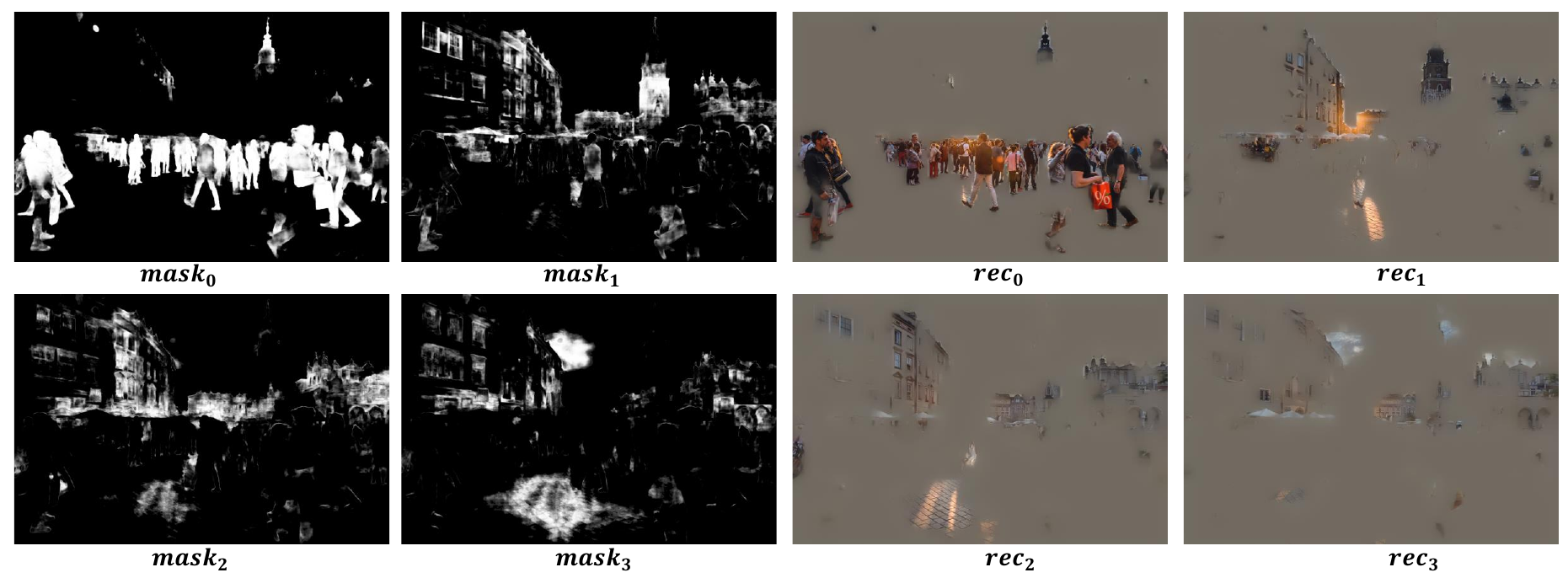}
    \caption{Object coding via applying the ROI masks into the latents. The entangled latents within channel dimensions represent different regions of the image.}
    \label{fig:hier_rec}
\end{figure*}



\begin{figure}
    \centering
    \includegraphics[width=0.99\linewidth]{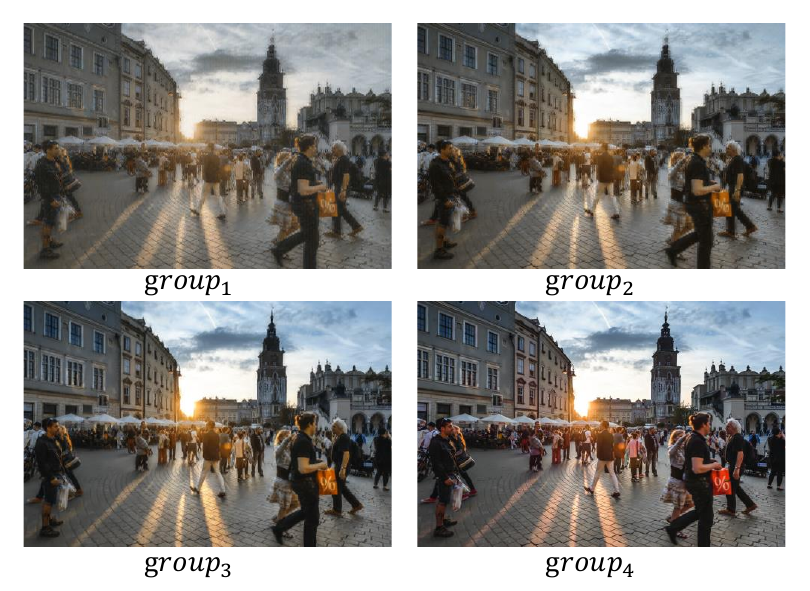}
    \caption{Progressive decoding with four groups. We restore the $group_{1}$ by setting latents of the latter groups $group_{2,3,4}$ to zeros, and $group_2$ by setting $group_{3,4}$ to zero, others analogically.}
    \label{fig:progressive_quant}
\end{figure}

\begin{figure}
    \centering
    \includegraphics[width=0.99\linewidth]{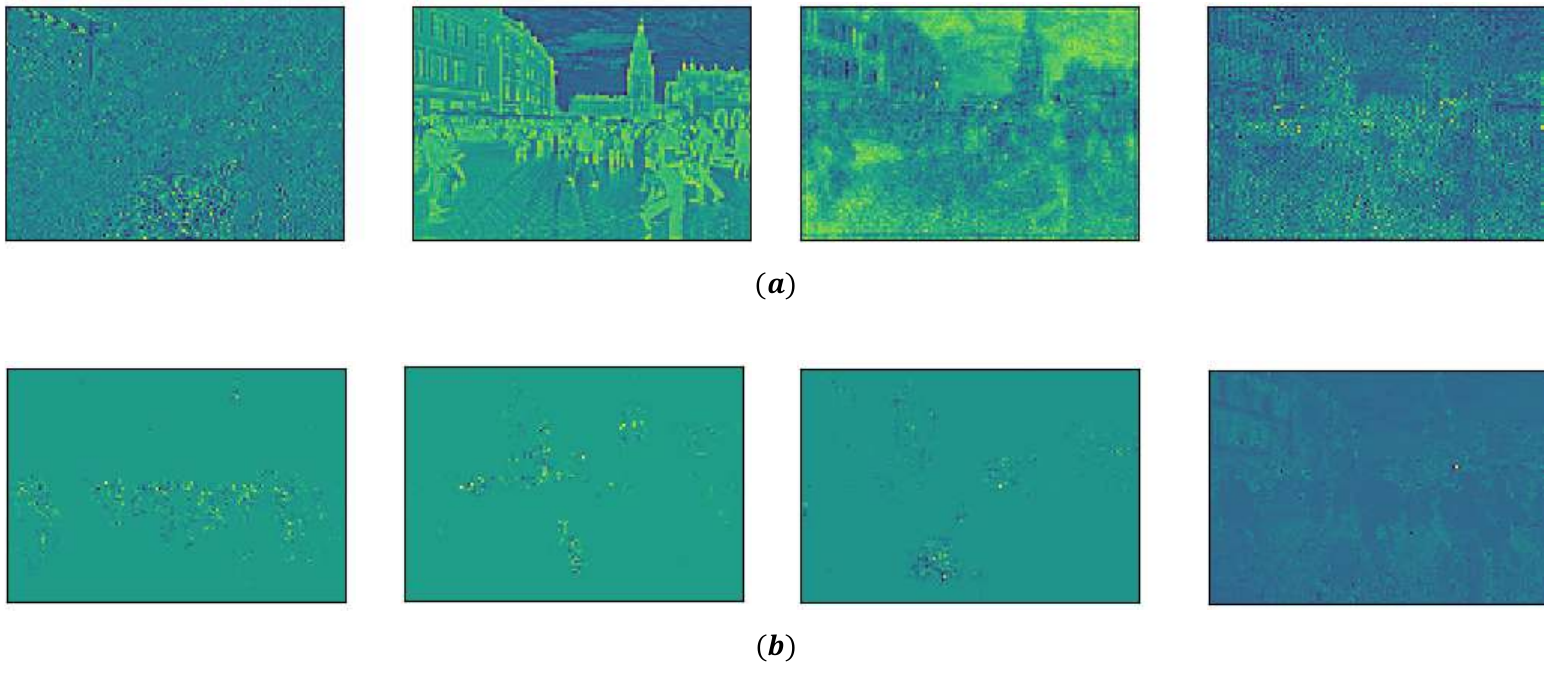}
    \caption{Visualization of latents. The above part $(a)$ is the latents without mask mapping, while $(b)$ shows results of mask mapping with the two-layer ROI. Latents of $(a)$ are region-agnostic while in $(b)$ the first half of channels correspond to the foreground and the second half represents the background.}
    \label{fig:hier_rec_latents}
\end{figure}

\begin{figure}
    \centering
    \includegraphics[width=0.9\linewidth]{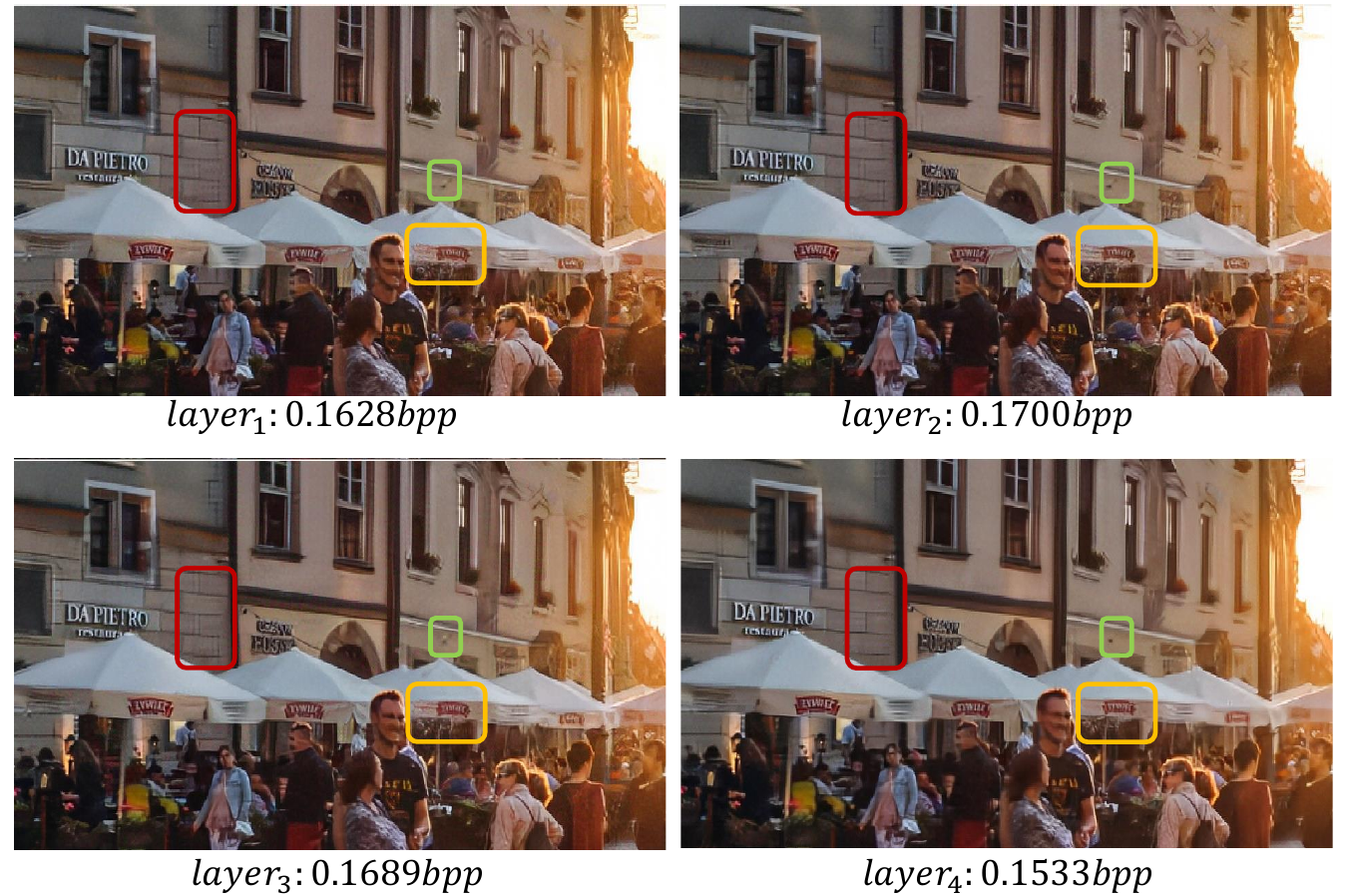}
    \caption{Ablation study for different layers of H-ROI. Red, yellow and green rectangles represent the regions lying out of the first foreground region $F_1$.}
    \label{fig:h_roi_adaptive}
\end{figure}

\section{Experiments}

\subsection{Training settings}
We split the training process into two stages: first, we use MSE loss to train the ELIC models and use the binary cross-entropy loss to train the salient detection network. Then, we fix the weights of the salient detection network and utilize H-ROI to train the framework of ELIC. For the salient detection network, we follow the training setting of PCSA~\cite{gu2020PCSA}. 

We use the subset of ImageNet~\cite{deng2009imagenet} with the number of 8000 in both stages to train ELIC. For the first stage we train 2000 epochs with a learning rate of 1e-4 and batch size of $2$ while for the second stage we train 300 epochs with a learning rate of 5e-5. Besides, we use learning rate decay with a ratio of $0.9$ after every 60 epochs at the second stage to avoid the overfitting caused by perceptual loss and adversarial loss. Meanwhile, we train each model with Adam optimizer.

For different bit rates we adjust parameter $\lambda$ in Eq.~\ref{rd_loss} from the set of $\{3, 8, 15, 200\} \times 10^{-4}$. We first use $\lambda = 200 \times 10^{-4} $ to train one model with an extremely high bit rate, around $1.0bpp$, and then we take it as the basis for other models. We choose the left parameters $\{3, 8, 15\} \times 10^{-4}$ to obtain the models with extremely low bit rates from $0.08bpp$ to $0.3bpp$ on Kodak dataset~\cite{franzen1999kodak}. We set the total channel number to $320$ in Fig.~\ref{fig:quant_layer2} for all models.

It takes about 30 hours with 8 GPUs (Tesla PG503-216) for the training of the first stage, while we only need 6 hours for the second stage. Since we utilize only one pre-trained model of ELIC and PCSA, the total training resource we use is limited. 

\subsection{Testing settings}
We utilize Kodak~\cite{franzen1999kodak} to evaluate the performance of the codec. To further demonstrate the effectiveness of our method, 30 images selected from the CLIC2022 ~\cite{clic2022} testing dataset are used and we randomly choose 30 pictures from CrowdHuman~\cite{shao2018crowdhuman} testing dataset with 5018 pictures, which consists of various resolutions and numerous scenarios with small faces or text. 




\subsection{Quantitative Results}\label{quan_index}
Since Zhang~\etal~\cite{unreasonable} demonstrates the unreasonable assessment for the objective metrics of visual quality, we utilize PSNR, MS-SSIM and LPIPS\cite{zhang2018unreasonable} for an evaluation more consistent with human eyes to verify the effectiveness of our method. PSNR and MS-SSIM represent fidelity and LPIPS aims to evaluate the realism of images. Thus, we combine these three metrics to measure the performance of codecs to consider both fidelity and reality.  

As shown in Tab.~\ref{tab:quantitative} and Fig.~\ref{fig:lpips}, our method achieves the lowest LPIPS among ELIC, HiFiC and BPG, while maintaining a PSNR close to BPG, and MS-SSIM superior to BPG. To be specific, we calculate LPIPS by PIQ~\footnote{https://github.com/photosynthesis-team/piq} with normalization in $[0, 1]$ with channels in BGR order. The lower LPIPS is, the better the reconstruction is. Besides we calculate the bit saving to BPG and HiFiC when they achieve the comparable LPIPS. When referring to Kodak dataset in Fig.~\ref{fig:lpips}, we obtain more than 50\% bits saving over BPG and ELIC, and more than 30\% over HiFiC. 


\subsection{Ablation study}
\textbf{Influence of the Number of Layers for Adaptive Quantization}. Fig.~\ref{fig:quant_layer} demonstrates the effectiveness of adaptive quantization with different channel groups. We set three types of adaptive quantization, each denoted as ${layer_2, layer_3, layer_4}$ compared to no adaptive quantization, denoted as ${layer_1}$, where ${layer_2}$ has two channel groups with $32, 288$ channels, ${layer_3}$ has three groups with $32, 64, 224$ channels and ${layer_4}$ has four groups with $32, 64, 72, 152$ channels. When we set more channel groups with larger $\epsilon_i$, the bit rate in Fig.~\ref{fig:quant_layer} decreases while the fidelity of reconstructed small faces and text is preserved.

For ${layer_4}$, we set the four groups with $\epsilon_1 = 0.5, \epsilon_2=0.4, \epsilon_3=0.3$ and $\epsilon_4=0.2$ respectively. The former groups quantize more values to the lower bounds while for the latter group, the opposite is true. Thus, adaptive quantization with channel groups enables progressive coding which improves the visual quality of internal decoding stages as shown in Fig.~\ref{fig:progressive_quant}. We restore the images with the former groups $group_{i}$ and set the latter groups $group_{j>i}$ to zeros. From $group_1$ to $group_4$, the reconstructed image has higher and higher fidelity of color and texture. Meanwhile, the reconstruction of $group_3$ is close to that of $group_4$ because of the low $\epsilon$ value of $group_4$. 

\textbf{Mask of H-ROI with Latents for Objective Coding}\label{Objective}: To demonstrate the effectiveness of H-ROI mask with PCSA, we further apply the mask of H-ROI on latents. Similar to Fig.~\ref{fig:quant_layer2}, we split the total $c_n$ channels into several groups, and each group is combined with mask by element-wise multiplication as shown in Eq.~\ref{eq:roi-latent}. 
We train and infer the network with masks generated by the PCSA network on $\hat y$. To be specific, 
we only visualize four channels to distinguish the latents with the same channel index $y_{0, i,j}, y_{60, i,j}, y_{120, i,j}, y_{240, i,j}$. Each latent in Fig.~\ref{fig:hier_rec_latents} (a) consists of all image contents differing only in frequency. While the latents in Fig.~\ref{fig:hier_rec_latents} (b)  are extracted from four different groups split with  $c_n$ channels, which distinguish the foreground with human and text, and the background represented by the latents of the right part of Fig.~\ref{fig:hier_rec_latents} (b). Thus, the latents without mask are region-agnostic and contain the global information generated by convolutional kernel. While the latents with masks are entangled and hierarchical representations can independently capture different regions, which is learned by leveraging the ROI mask. 

We further utilize the multi-layer ROI masks to reconstruct different regions of images as shown in Fig.~\ref{fig:hier_rec} with $mask_{0,1,2,3}$ to obtain the reconstruction $rec_{0,1,2,3}$.  When we restore one certain region of the image, we set all other channels to zero. Thus the channels of latents can generate corresponding objects for different regions of images. 

It is promising to apply the disentangled latents to object detection or segmentation with only the related regions encoded and without other redundant information. Thus,  image or video coding for machine can be processed at a very low bit rate. However, coding for machine is beyond the scope of this paper. We leave it as future work. 

\textbf{Influence of the Number of Layers for H-ROI}. In our experiments, we utilize four layers as our default settings. We do not consider using more layers because as the number of layers increases, fewer salient objects are detected. As shown in Fig.~\ref{fig:hier_rec}, $rec_{2, 3}$ contain fewer regions, and there is no need to increase computation complexity when the gain is negligible.

We compare the reconstructions of Hierarchical-ROI with different numbers of layers with no foreground and ${1,2,3}$ layers of foreground and one background. First, we reconstruct them all around $0.16bpp$, and then adjust the number of foregrounds. We obtain images with different visual qualities as shown in Fig.~\ref{fig:h_roi_adaptive}. Regions emphasized are marked with red, green and yellow rectangles. For red and yellow regions the reconstruction with $layer_4$ is of higher fidelity without fake textures generated by the adversarial neural network, such as the boundary of brick in the red rectangle and the unnatural color shift noise in the yellow rectangle. Besides, the point in the green rectangle of $layer_4$ is of higher fidelity compared with others.

\section{Conclusion}
We propose Hierarchical-ROI to detect hierarchical salient regions and maintain image fidelity by applying the MSE loss function with decreasing Lagrange multipliers. Then, we introduce adaptive quantization with non-linear mapping to further reduce the bit rate without compromising visual quality. We maintain the visual quality with bit savings of more than \textbf{30\%} compared with HiFiC and more than \textbf{50\%} compared with BPG with regard to LPIPS.

We discuss the concept of objective coding in Sec.~\ref{Objective}  by combining the mask of H-ROI and latents. The latent is content-aware and we can reconstruct each region independently by setting other channels to zero. It is worth studying for online conferences or meetings by minimizing the bit cost of background. 
And it can facilitate downstream tasks like object detection since we just need to encode related regions with an extremely low bit rate. 


\balance

{\small
\bibliographystyle{ieee_fullname}
\bibliography{PaperForReview}
}

\end{document}


\title{Supplementary \\
\normalsize{Super-High-Fidelity Image Compression via Hierarchical-ROI and Adaptive Quantization} }
\author{Jixiang Luo\\
Sensetime Research\\
{\tt\small jixiangluo85@gmail.com}
\and
Yan Wang\\
Tsinghua University\\
{\tt\small wangyan@air.tsinghua.edu.cn}
\and
Hongwei Qin\\
Sensetime Research\\
{\tt\small qinhongwei@sensetime.com}
}
\maketitle

\section{Dataset Detail}
To reproduce our results, we list the subset of CrowdHuman testing dataset as shown in Tab.~\ref{tab:crowdhuamn}, where $w, h$ are the width and height, and resolution varies from $615 \times 461$ to $4214 \times 2730$. Besides, it consists of numerous people with various gestures and facial expression, which is more close to real situation.  Different from Kodak and CLIC2002 testing dataset,  its image format is JPEG and its quality is 
relatively inferior because of  information loss and JPEG compression noise. However JPEG is the most popular format for transmission or storage, thus  our experiment setting is more practical for industrial application. 

\section{Qualitative Results }
We compare the original image, reconstructions from BPG, ELIC, HiFiC and ours(H-ROI) with regard to visual quality along with bpp, objective metrics PSNR and MS-SSIM, and subjective index LPIPS.  We select images from Kodak dataset, CLIC2022 testing dataset and a subset of CrowdHuman testing dataset. The last row is the detail for local area of  BPG, ELIC, HiFiC and H-ROI. 

\begin{figure}
    \centering
    \includegraphics[width=.9\linewidth]{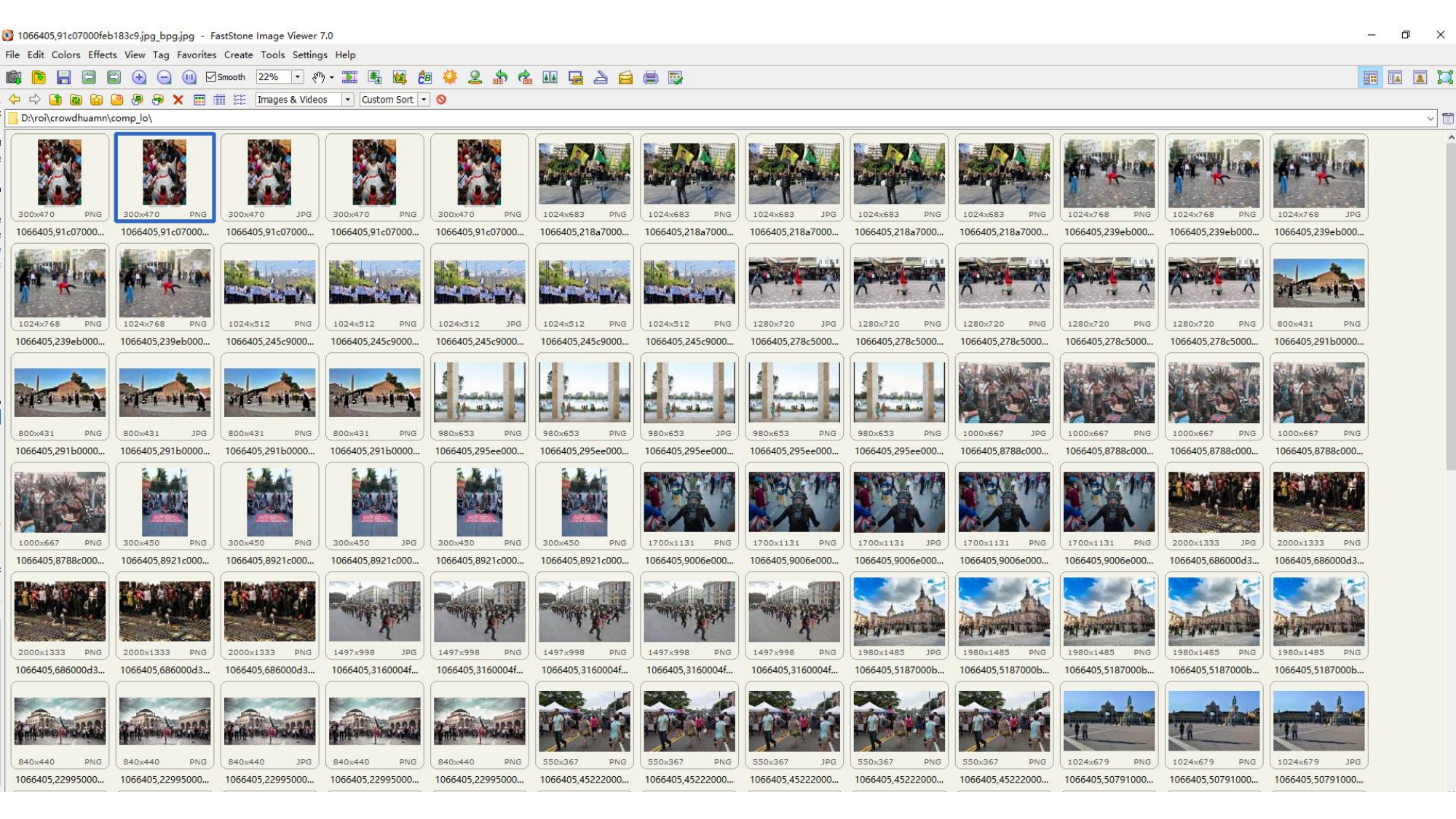}
    \caption{The overall of folder in FastStone.}
    \label{fig:faststone1}
\end{figure}

\begin{figure}
    \centering
    \includegraphics[width=.9\linewidth]{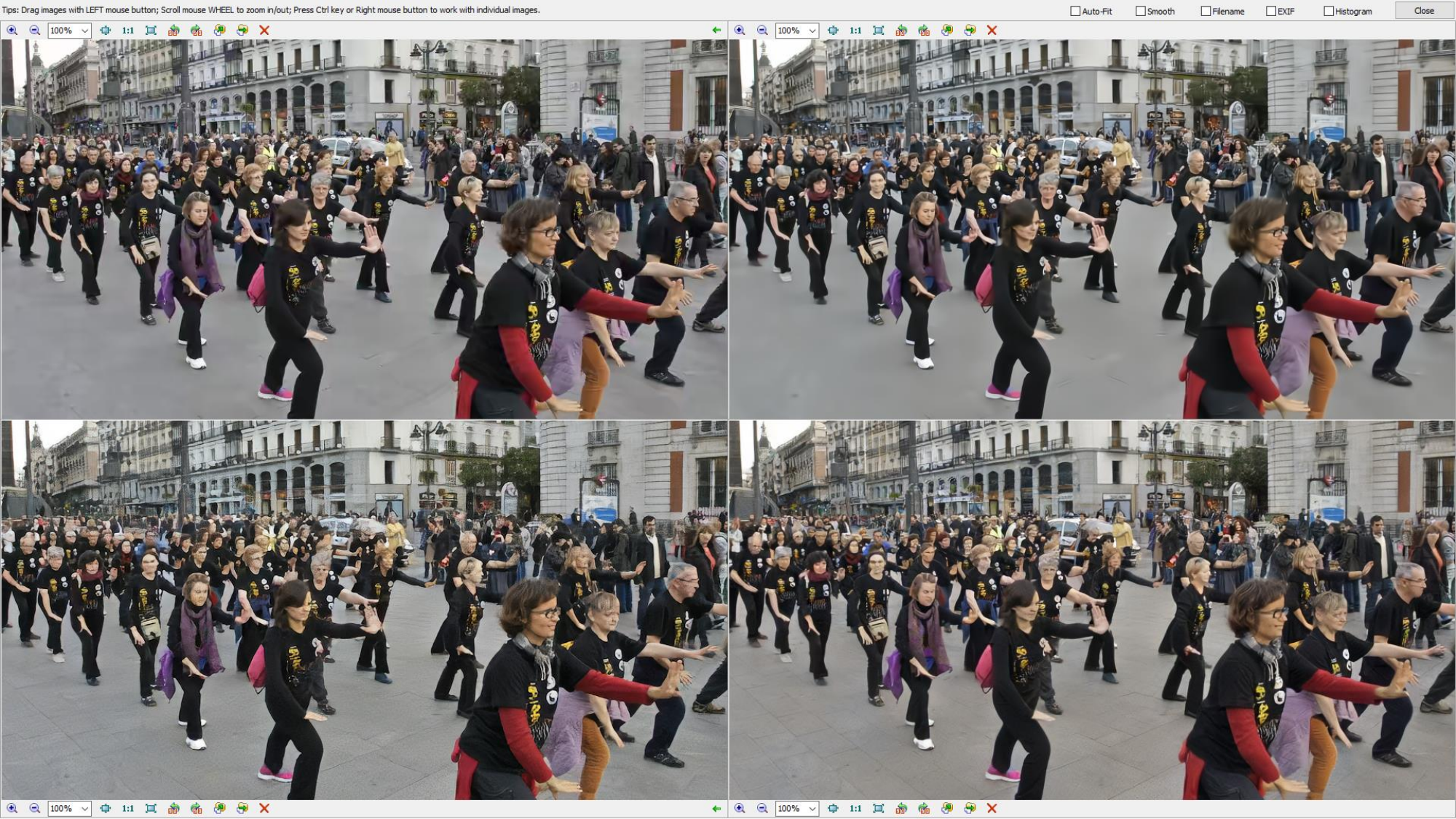}
    \caption{Compare pictures at the same screen with maximal number of 4 in FastStone. }
    \label{fig:faststone2}
\end{figure}

With the combination of PSNR, MS-SSIM, LPIPS, and the visual quality from human eyes using FastStone~\footnote{https://www.faststone.org/}, the reconstructions from our methods are more pleasurable with regard to other methods. 
Fig.~\ref{fig:faststone1} and Fig.~\ref{fig:faststone2} depict the working diagram of FastStone, and it can zoom the picture in or out for all pictures with the same scale. Thus we can evaluate the smallest region such as $20 \times 20$ for width and height to distinguish which details have been missing or which structure is more complete.  Besides, for the following visual comparisons, the last row marked with BPG,  ELIC, HiFiC, H-ROI is captured by  FastStone for  a small region. Moreover, we evaluate the image from global  to local perspective with FastStone.

\begin{table}[]
\caption{Subset of CrowdHuman}\label{tab:crowdhuamn}
\centering
\begin{tabular}{c|c|c}
\hline
name                     & w    & h    \\ \hline
1066405,218a700048baf6a5 & 1024 & 683  \\ \hline
1066405,22995000cde69f78 & 840  & 440  \\ \hline
1066405,239eb000e5646ae4 & 1024 & 768  \\ \hline
1066405,245c90002f8426d4 & 1024 & 512  \\ \hline
1066405,278c500023fc022b & 1280 & 720  \\ \hline
1066405,291b00006e60dcd4 & 800  & 431  \\ \hline
1066405,295ee0009ddbbec8 & 980  & 653  \\ \hline
1066405,3160004fdd79b1   & 1497 & 998  \\ \hline
1066405,45222000bb360294 & 550  & 367  \\ \hline
1066405,50791000dde5fe22 & 1024 & 679  \\ \hline
1066405,5187000bbd15ae1  & 1980 & 1485 \\ \hline
1066405,686000d3439407   & 2000 & 1333 \\ \hline
1066405,8788c000723bb59d & 1000 & 667  \\ \hline
1066405,8921c00054293f10 & 300  & 450  \\ \hline
1066405,9006e0007cd7a2ab & 1700 & 1131 \\ \hline
1066405,91c07000feb183c9 & 300  & 470  \\ \hline
1066405,c264200072c9b890 & 640  & 458  \\ \hline
1066405,caf4c00088060824 & 2000 & 1308 \\ \hline
1066405,cbb81000e1b4a4a6 & 2400 & 1600 \\ \hline
1066405,cda8500076abec04 & 2048 & 1174 \\ \hline
1066405,ce1710005ce68455 & 3180 & 2204 \\ \hline
1066405,ceb8c000c098b409 & 3264 & 2448 \\ \hline
1066405,cedc80008a06a63c & 3238 & 2850 \\ \hline
1066405,ceedc000dc09b8e1 & 1024 & 683  \\ \hline
1066405,d050008daf0bde   & 900  & 675  \\ \hline
1066405,d583000a00e73e2  & 4214 & 2730 \\ \hline
1066405,db19000e046d964  & 800  & 600  \\ \hline
1066405,dd9b000d31071df  & 615  & 461  \\ \hline
1066405,e77810002028a57e & 1500 & 1125 \\ \hline
1066405,e9ec100081152e8c & 1600 & 2397 \\ \hline
\end{tabular}
\end{table}

\begin{figure}
    \centering
    \includegraphics[width=0.49\linewidth]{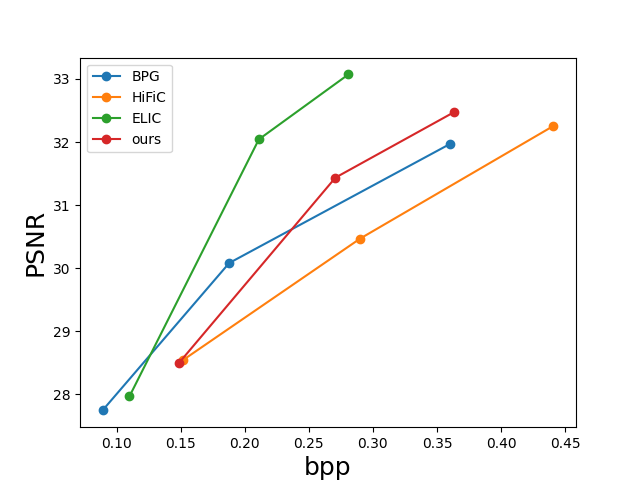}
    \includegraphics[width=0.49\linewidth]{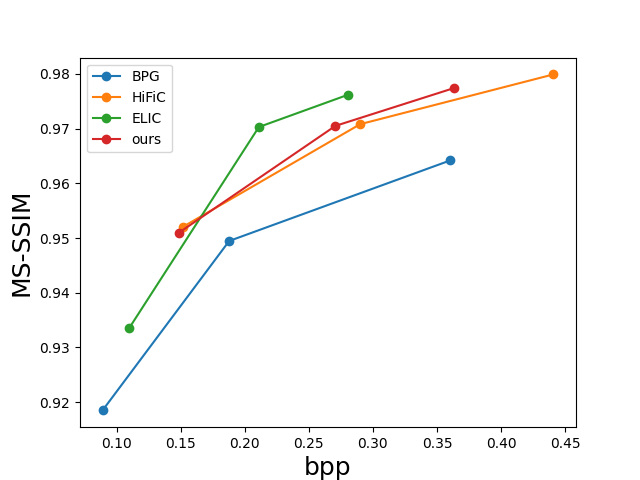}
    
    \includegraphics[width=0.49\linewidth]{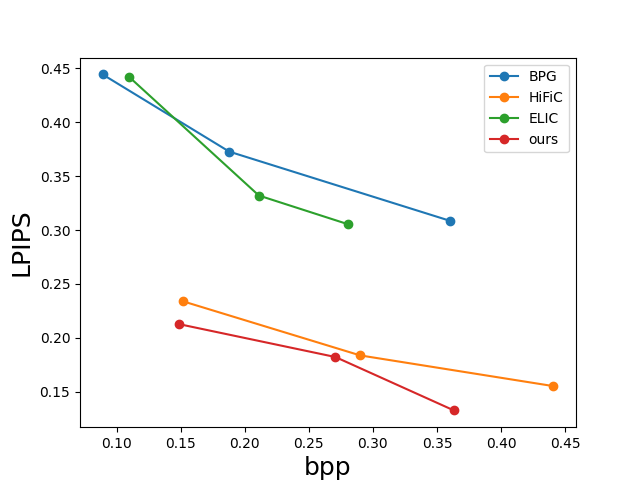}
    \caption{Performance on PSNR, MS-SSIM and LPIPS of our method, HiFiC, ELIC and BPG on CLIC2022 testing dataset.}
    \label{fig:clic30-lpips}
\end{figure}

\begin{figure}
    \centering
    \includegraphics[width=0.49\linewidth]{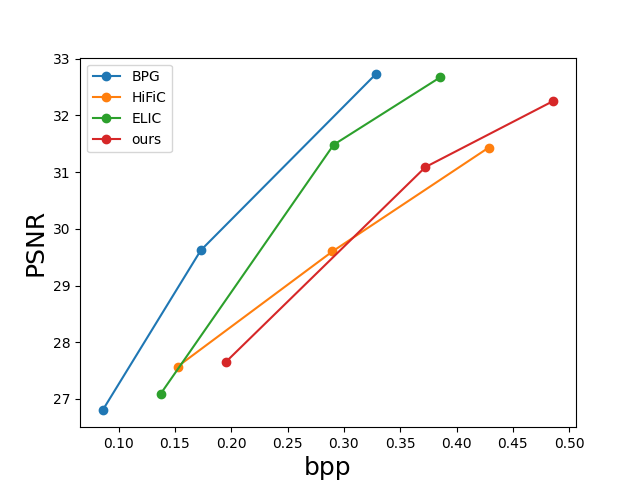}
    \includegraphics[width=0.49\linewidth]{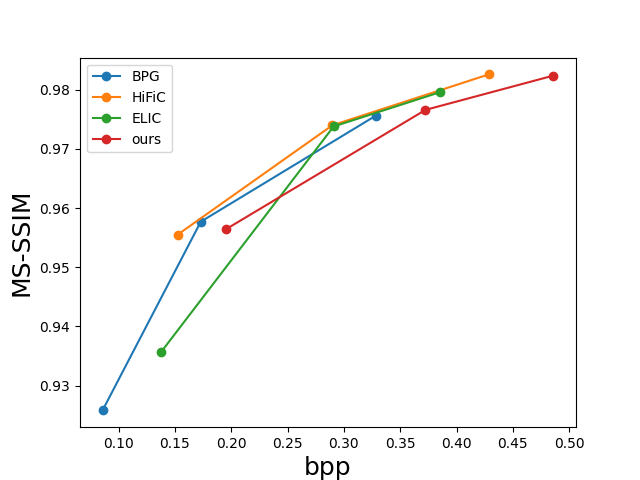}
    
    \includegraphics[width=0.49\linewidth]{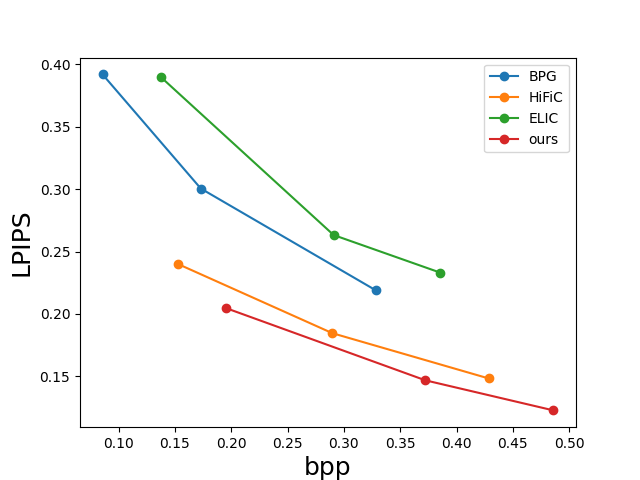}
    \caption{Performance on PSNR, MS-SSIM and LPIPS of our method, HiFiC, ELIC and BPG on subset of CrowdHuman testing dataset.}
    \label{fig:crowdhuman-lpips}
\end{figure}

In Fig.~\ref{fig:clic30-lpips} and Fig.~\ref{fig:crowdhuman-lpips}, we achieve the lowest LPIPS and comparable MS-SSIM and PNSR for HiFiC in CLIC2022 testing dataset and the subset of CrowdHuman testing dataset. Thus our methods occupy the merit of generalization from small resolution $512 \times 768$ or $615 \times 461$ to larger resolution $4124 \times 2730$. 
Our method has the smallest bit consumption, but achieves comparable or better visual quality for face, text or other textures. Besides, we maintain the detail of other regions as shown in the following visualization.

\section{Object Codec}
\begin{figure}
    \centering
    \includegraphics[width=.9\linewidth]{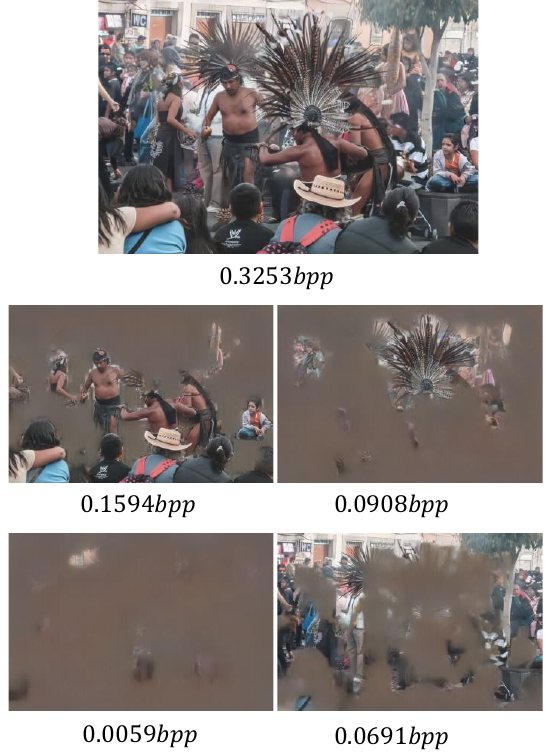}
    \caption{Reconstruction for objective coding}
    \label{fig:od1}
\end{figure}

\begin{figure}
    \centering
    \includegraphics[width=.99\linewidth]{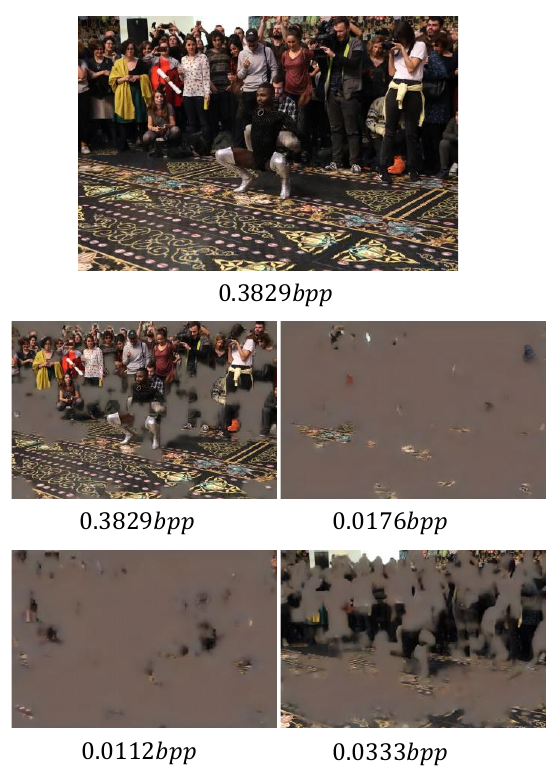}
    \caption{Reconstruction for objective coding}
    \label{fig:od2}
\end{figure}

\begin{figure}
    \centering
    \includegraphics[width=.99\linewidth]{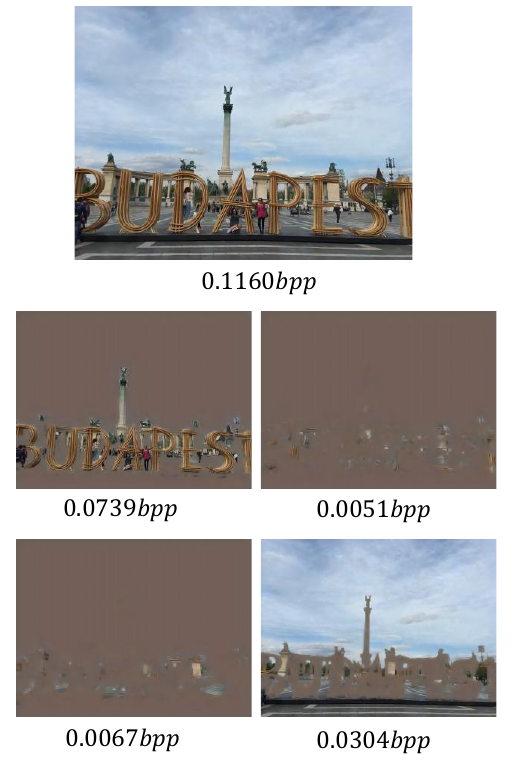}
    \caption{Reconstruction for objective coding}
    \label{fig:od3}
\end{figure}

We further study the objective coding via mask for latents, then we can reconstruct the region of interests. 
We visualize more reconstruction from only a part of latents corresponding to certain objects. The top row in Fig.~\ref{fig:od1}, Fig.~\ref{fig:od2} and Fig.~\ref{fig:od3} is the complete reconstruction. The left rows are the partial reconstruction from the part of channels. To be specific, we calculate the bpp of each part, and the sum of bpp in part channels is equal to the total bpp.

\begin{figure*}
    \centering
    \includegraphics[height=.99\textheight]{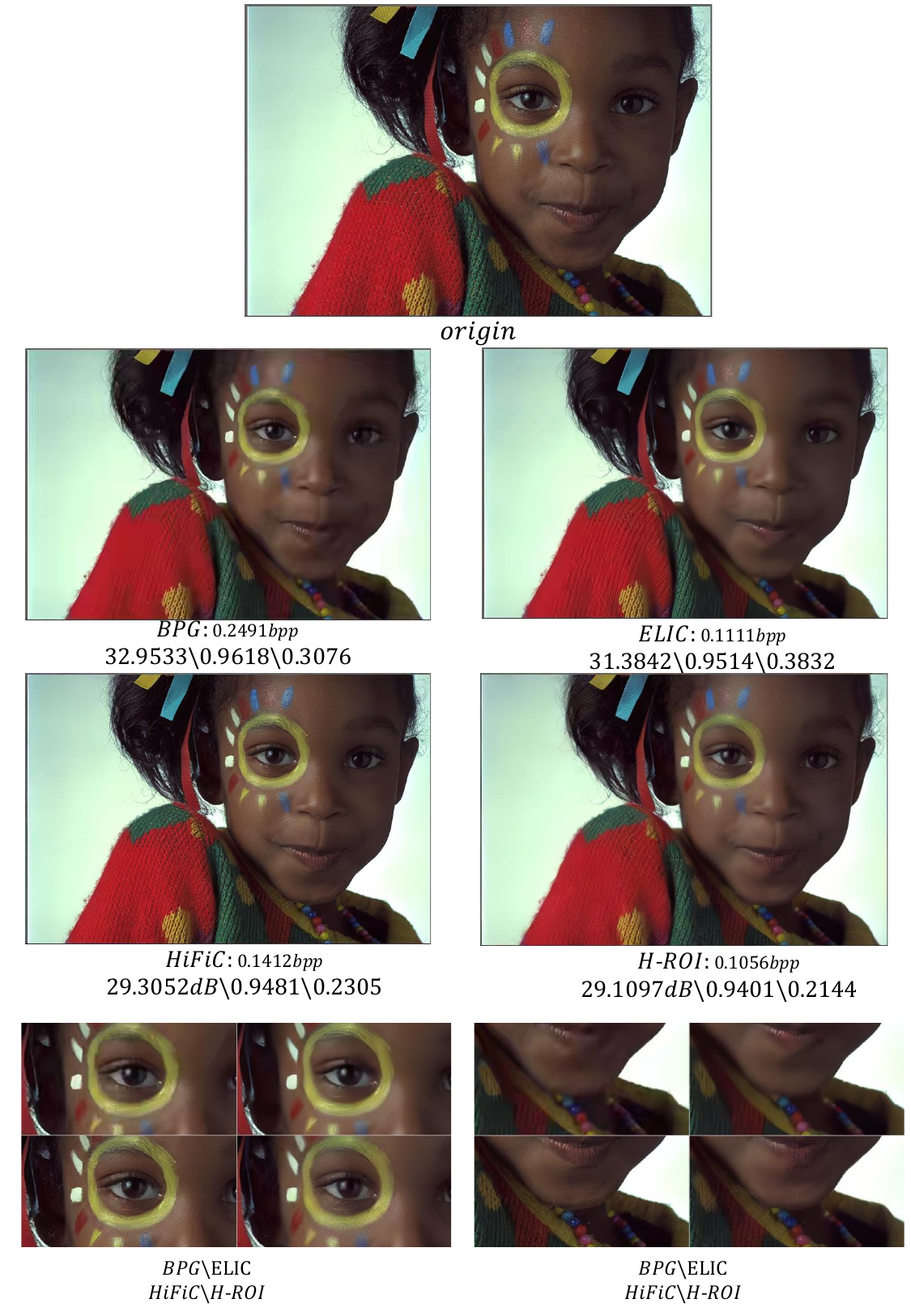}
\end{figure*}

\begin{figure*}
    \centering
    \includegraphics[height=.99\textheight]{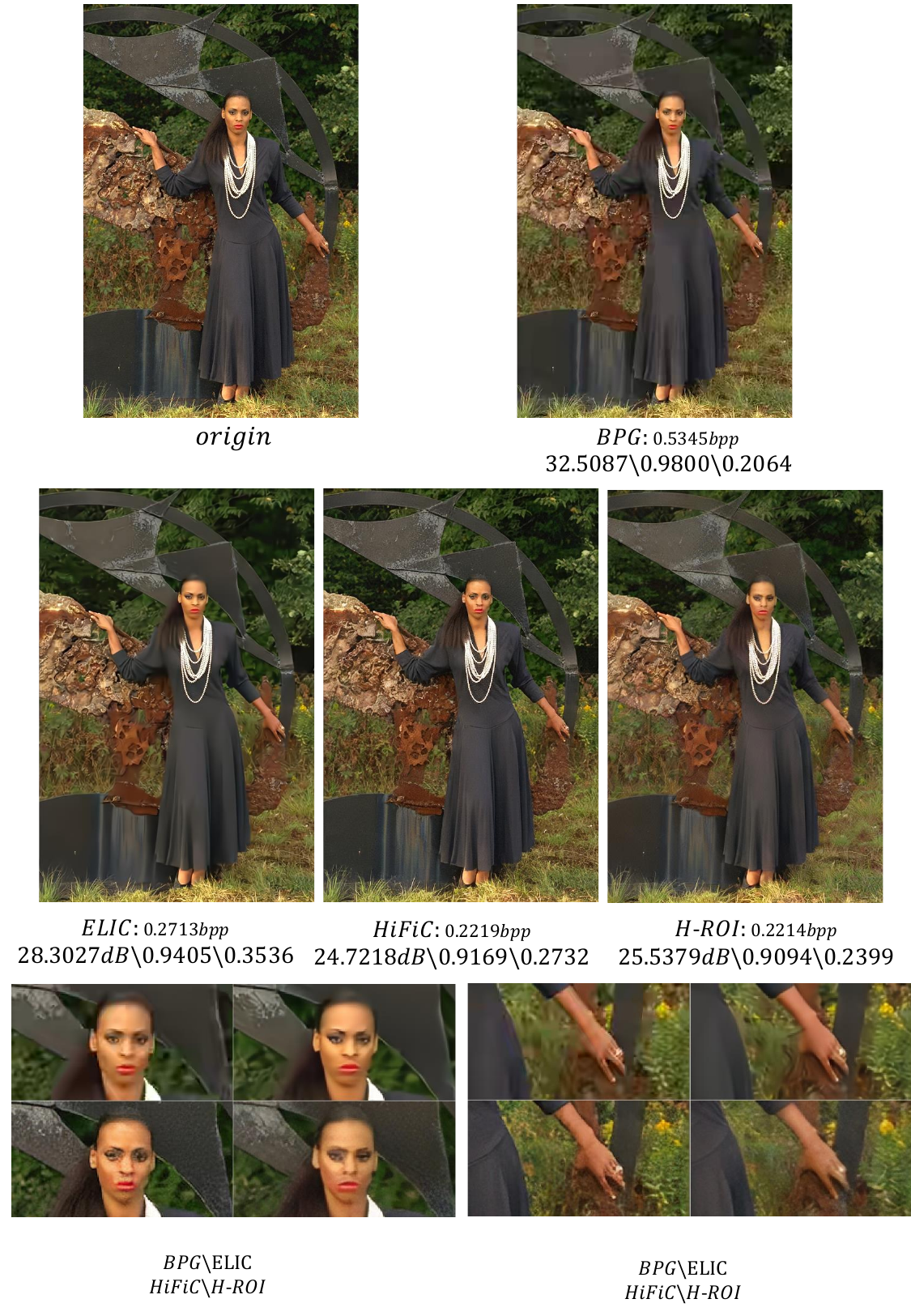}
\end{figure*}

\begin{figure*}
    \centering
    \includegraphics[height=.99\textheight]{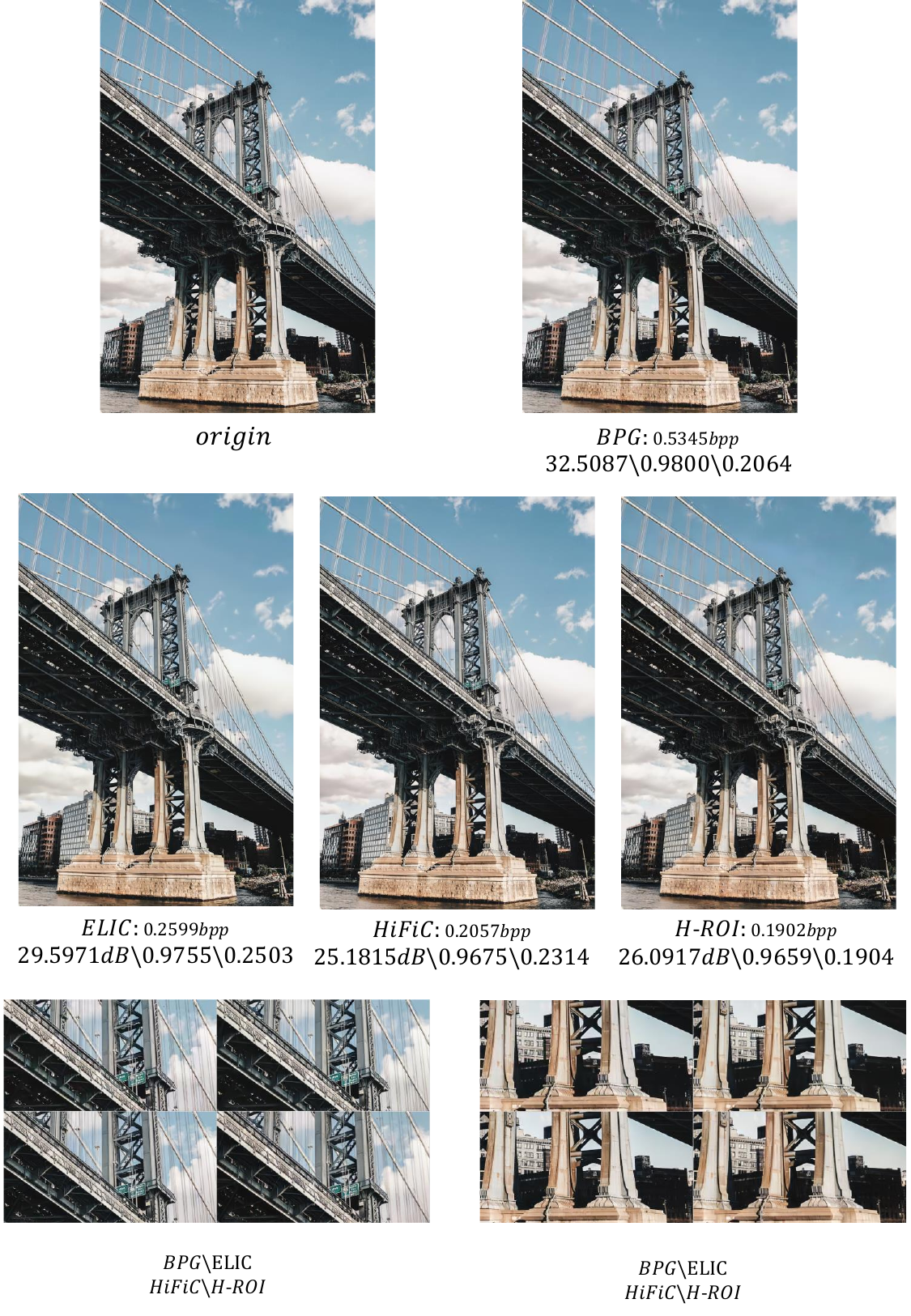}
\end{figure*}

\begin{figure*}
    \centering
    \includegraphics[height=.99\textheight]{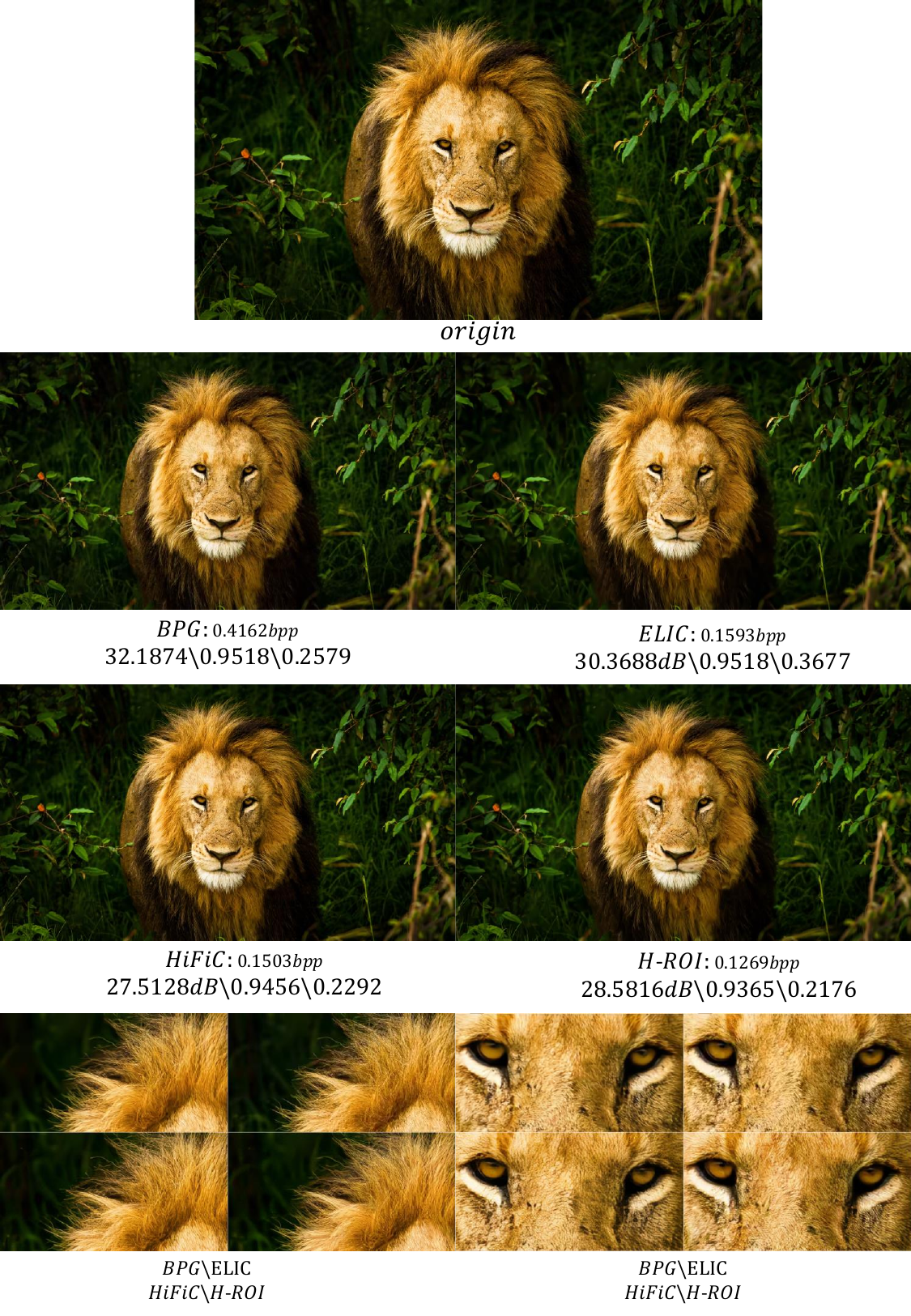}
\end{figure*}

\begin{figure*}
    \centering
    \includegraphics[height=.99\textheight]{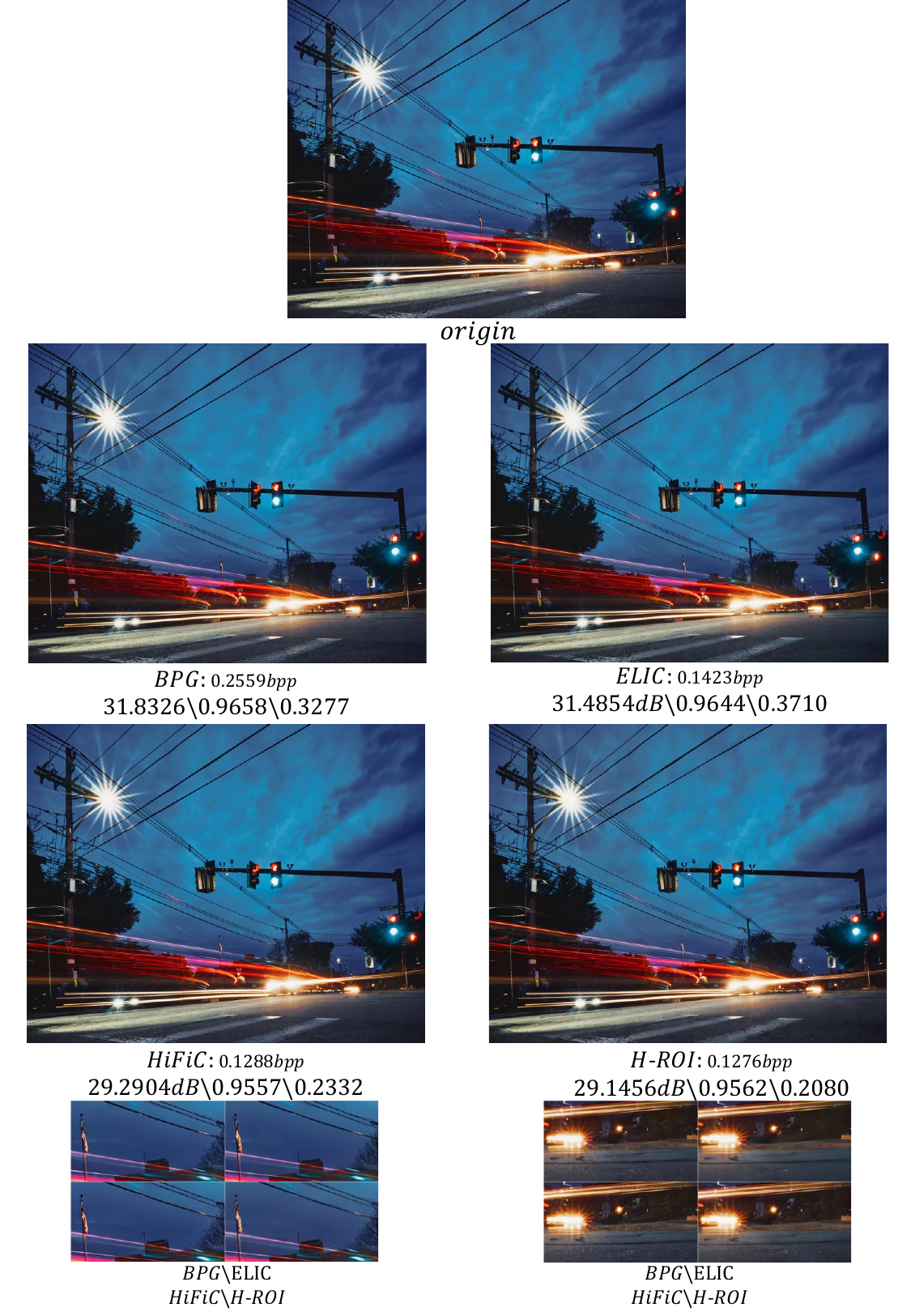}
\end{figure*}

\begin{figure*}
    \centering
    \includegraphics[height=.99\textheight]{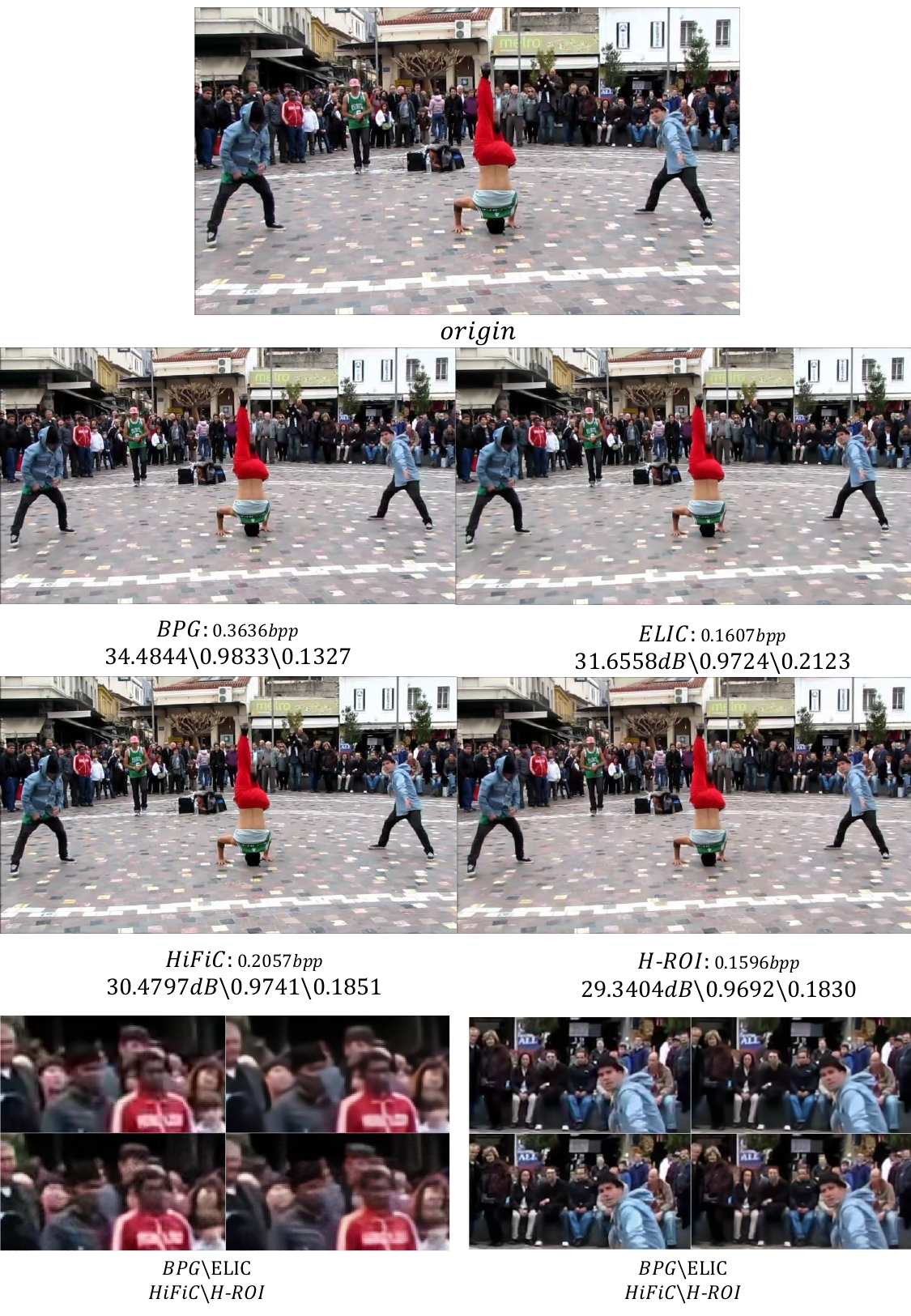}
\end{figure*}

\begin{figure*}
    \centering
    \includegraphics[height=.99\textheight]{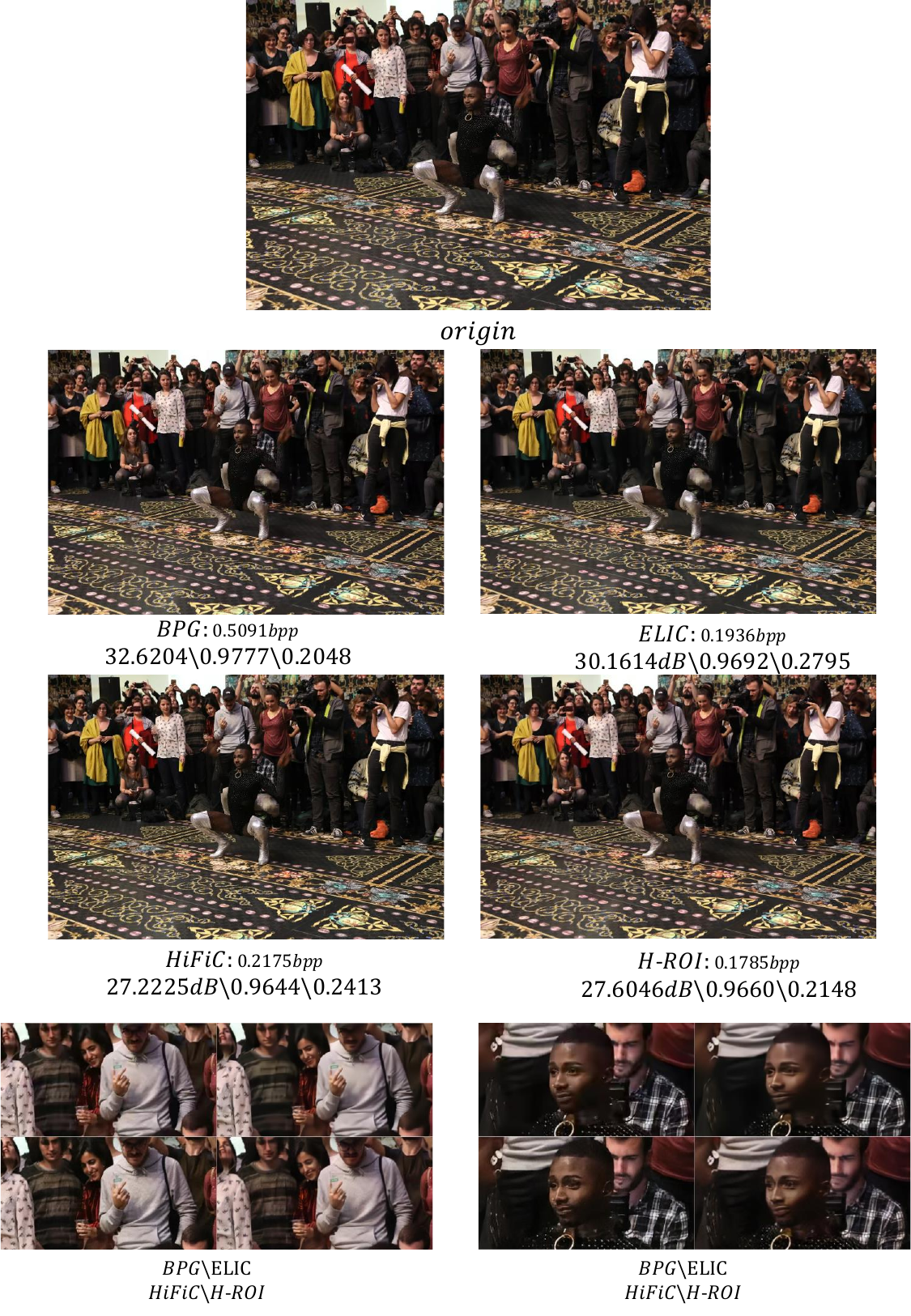}
\end{figure*}

\begin{figure*}
    \centering
    \includegraphics[height=.99\textheight]{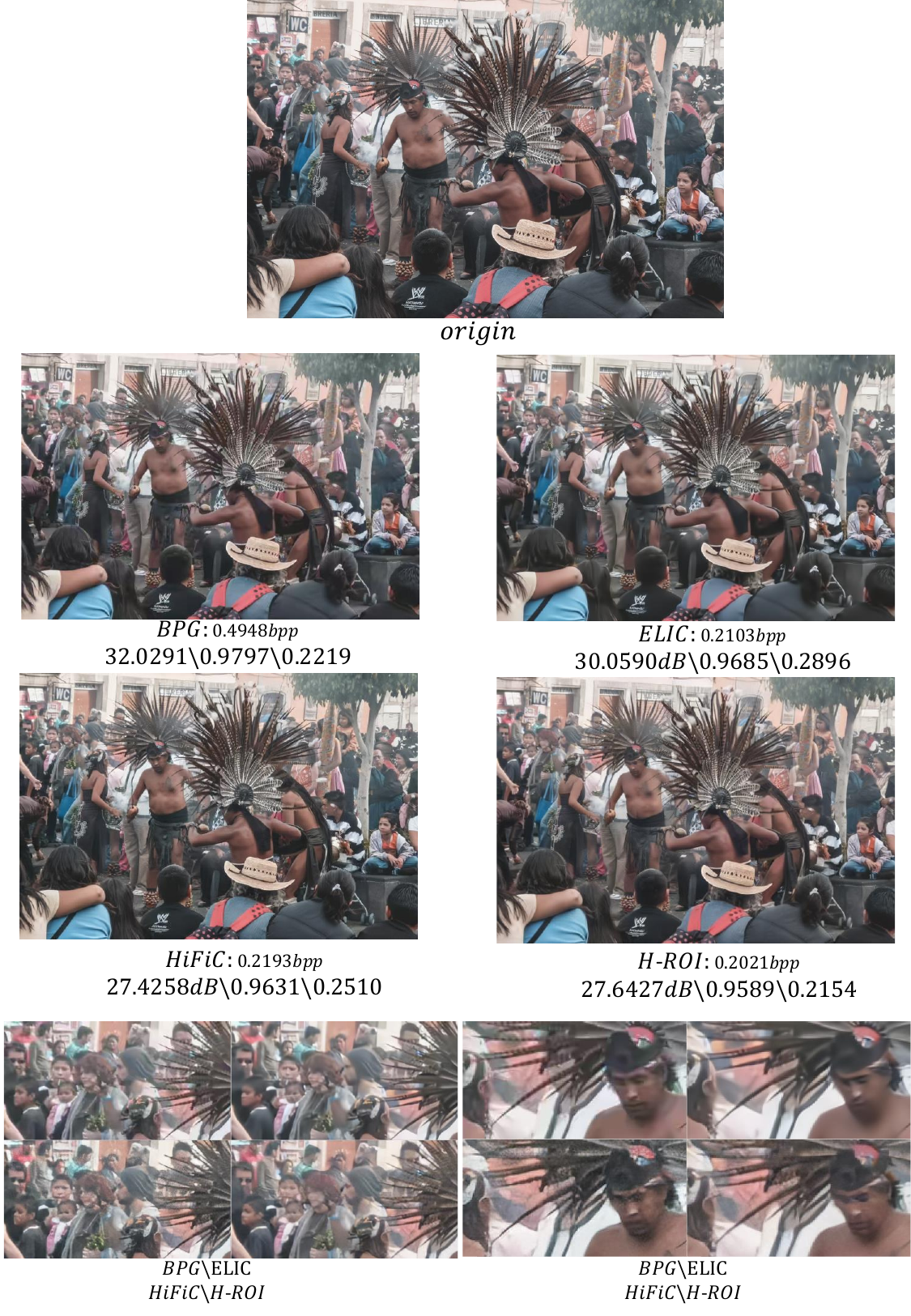}
\end{figure*}

\begin{figure*}
    \centering
    \includegraphics[height=.99\textheight]{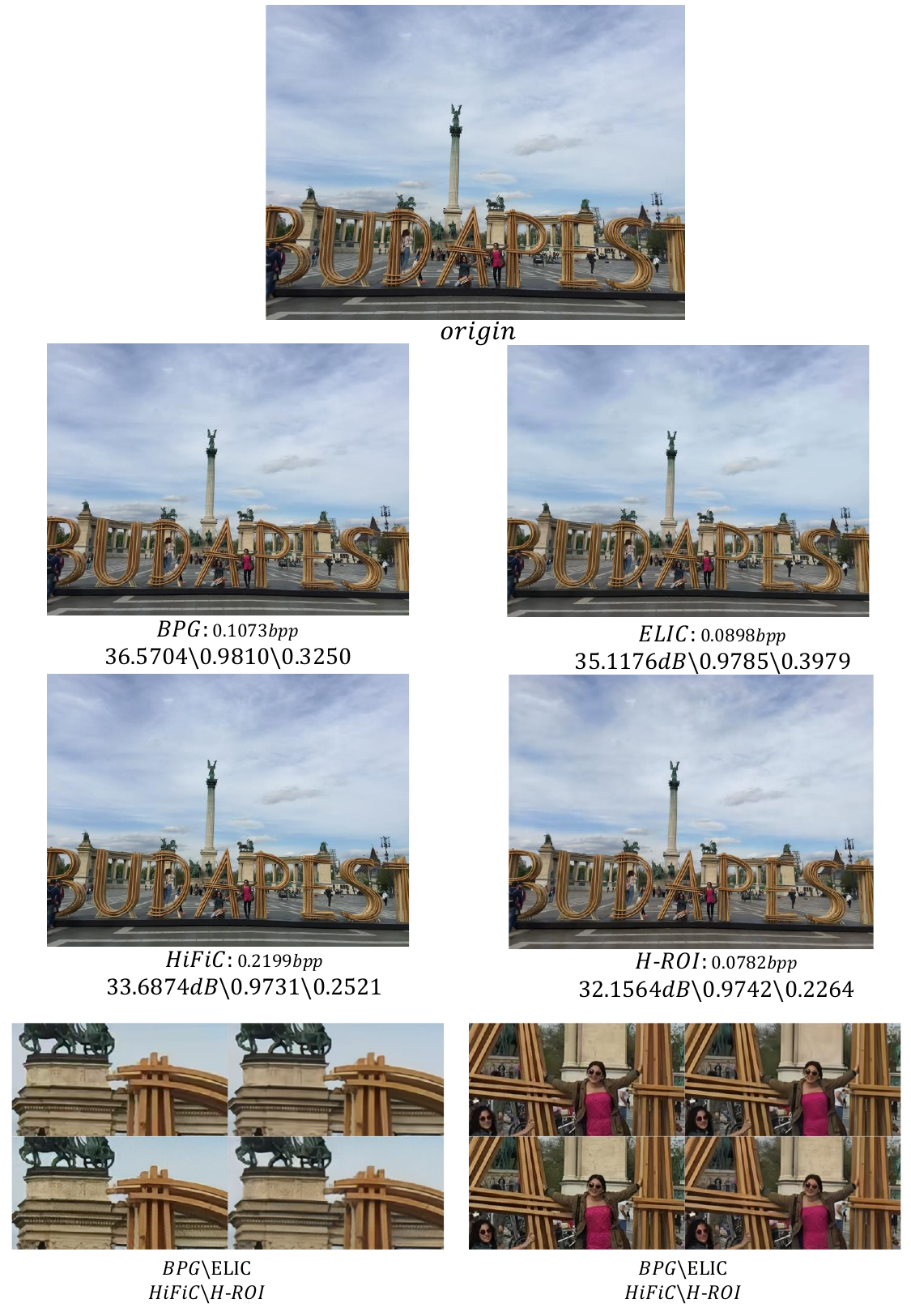}
\end{figure*}

\begin{figure*}
    \centering
    \includegraphics[height=.99\textheight]{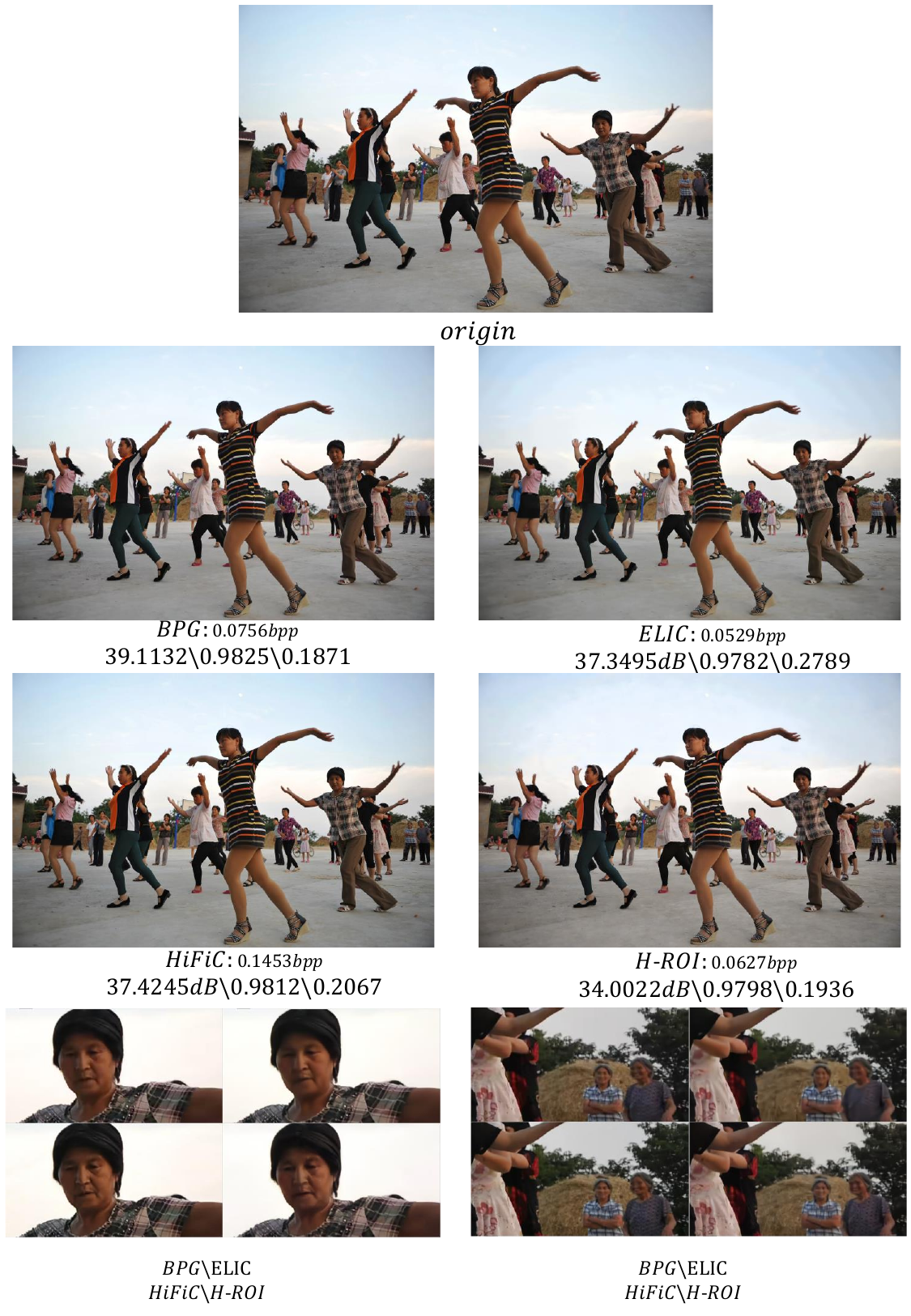}
\end{figure*}